\documentclass[11pt]{article}

\usepackage[T1]{fontenc}
\usepackage{amsmath,amssymb, graphicx, booktabs, hyperref, fancyhdr, titlesec, float, listings, xcolor, amsthm, array}
\usepackage{mathptmx} 
\usepackage{setspace}
\usepackage{enumitem}
\usepackage{algorithm}
\usepackage{algpseudocode}
\usepackage{adjustbox}
\usepackage{tablefootnote}
\usepackage{multirow}
\usepackage{longtable}
\usepackage{lipsum}
\usepackage{subcaption}
\usepackage{makecell}
\usepackage{soul}

\allowdisplaybreaks

\usepackage[margin=1in]{geometry}
\definecolor{codegreen}{rgb}{0,0.6,0}
\definecolor{codegray}{rgb}{0.5,0.5,0.5}
\definecolor{codepurple}{rgb}{0.58,0,0.82}
\definecolor{backcolour}{rgb}{0.95,0.95,0.92}

\hypersetup{
    colorlinks=true,        
    linkcolor=blue,         
    citecolor=blue,          
    urlcolor=blue           
}

\lstdefinestyle{mystyle}{
    backgroundcolor=\color{backcolour},   
    commentstyle=\color{codegreen},
    keywordstyle=\color{magenta},
    numberstyle=\tiny\color{codegray},
    stringstyle=\color{codepurple},
    basicstyle=\ttfamily\small,
    breakatwhitespace=false,         
    breaklines=true,                 
    captionpos=b,                    
    keepspaces=true,                 
    numbers=left,                    
    numbersep=5pt,                  
    showspaces=false,                
    showstringspaces=false,
    showtabs=false,                  
    tabsize=2
}

\title{An Augmented Rating System for Test Cricket:\\ adapting the Glicko rating system}

\author{
Rhitankar Bandyopadhyay$^{1}$, \ Diganta Mukherjee$^{2}$ \\[0.6em]
\normalsize $^{1}$University of Florida, Florida, USA. Email: \href{mailto:r.bandyopadhyay@ufl.edu}{r.bandyopadhyay@ufl.edu} \\
\normalsize $^{2}$Indian Statistical Institute, Kolkata, India. Email: \href{mailto:diganta@isical.ac.in}{diganta@isical.ac.in}
}

\date{}

\begin{document}

\maketitle

\begin{abstract}
\noindent The International Cricket Council's (ICC) rating system for Test cricket updates a team's rating from match and series results alone. It makes no allowance for the precision of a rating, nor for contextual factors such as home advantage and the outcome of the toss. This study develops an enhanced rating framework by adapting the Glicko rating system so that these influences enter a probabilistically grounded expected score. The rating scale is recalibrated for the dynamics of Test cricket, and the home-ground and toss effects enter as data-driven covariates whose statistical significance and relative weights are estimated directly from match data: playing at home is worth an advantage of about $13$ rating points and winning the toss about $8$, and the two effects combine additively, with no statistically significant interaction. Applied to the two completed World Test Championship cycles (2021-23 and 2023-25), the model correctly predicts $77.6\%$ of decisive matches in the first cycle. Benchmarked against the standard Elo and the unmodified Glicko systems, it matches the better of their predictive accuracies and produces considerably better calibrated win probabilities, with a lower Brier score and log loss in both cycles. A resampling test over $1000$ permutations of each cycle's match sequence shows every team's final rating to be invariant to the order in which the fixtures are played. We present this as robustness to match ordering and distinguish it from fairness with respect to the fixture list, which the unbalanced Test calendar does not allow us to assess. The resulting ordering of teams agrees very closely with the ICC's (Spearman rank correlation $0.979$ and $0.983$ at the end of the two cycles), so the improvement we demonstrate is not a different ranking but what accompanies it: a calibrated win probability for every match, an explicit rating deviation on every rating and contextual adjustments whose magnitude is estimated, none of which the current ICC scheme supplies. \\

\end{abstract}

\thispagestyle{empty}

\vspace{1.2cm}

\setcounter{page}{1}
\section{Introduction}\label{introduction}

Test cricket started way back in 1877 but it lacked a unified global competition structure until the introduction of the ICC World Test Championship in 2019. We discuss below several shortcomings of the rating system currently used by the International Cricket Council (ICC) which limit its ability to evaluate team performance consistently and transparently. Test cricket, being the longest and supposedly most complex format of the game, is influenced by a number of factors that the existing system does not account for. These include home ground advantage, the impact of winning the toss ahead of a match and the varying quality of opposition teams. Such factors matter a great deal to the outcome of a match and ought to form part of any system that sets out to assess teams on their performance. The unbalanced nature of the current schedule, where not all teams play against each other under similar conditions or the same number of times, compounds these limitations. We address them by developing a more robust and comprehensive rating framework for Test cricket, drawing inspiration from the Glicko rating system (\cite{glickman1995glicko}).

Probabilistically grounded rating systems for paired competition originate with the method of paired comparisons of Bradley and Terry (\cite{bradley1952rank}), whose logistic model of one competitor's dominance over another underlies most modern schemes. Arpad Elo later introduced a practical rating system for chess (\cite{elo1978rating}), updating each player's rating in proportion to the difference between the realised and expected result. The Glicko rating system was devised by \cite{glickman1995glicko} to remedy a specific deficiency of Elo, namely its neglect of the reliability of a rating. Glicko augments each rating with a rating deviation that quantifies its uncertainty and governs how much a result is allowed to move it. The technical development and derivation of the system, as a dynamic paired comparison model estimated by approximate Bayesian updating, are given by \cite{glickman1999parameter}. \cite{glickman2001dynamic} subsequently extends the model to stochastically varying variances, the basis of the successor system Glicko-2, for which \cite{glickman2012example} provides a worked example. A related uncertainty-aware, Bayesian generalisation of Elo is TrueSkill developed by \cite{herbrich2006trueskill}, which additionally accommodates draws and multi-player contests. Glicko and Glicko-2 are used by the Free Internet Chess Server, by Chess.com and Lichess and by a range of other online gaming platforms. The official United States Chess Federation ratings, by contrast, are computed by a separate Elo-based formula and are not directly based on Glicko ratings.

Rating systems for paired competition have been extended in several sports to accommodate contextual structure beyond relative strength. For soccer, \cite{dewenter2003raising} examines the effect of scoring rule changes, \cite{marek2020comparison} study home advantage in football and rugby, and \cite{kovalchik2020extension} extends the Elo framework to incorporate the margin of victory in tennis. These papers share a template: augment a base rating system with sport-specific covariates. We follow it here for Test cricket.

Statistical modelling of cricket outcomes and team strength has itself received considerable attention, although rarely through an uncertainty-aware rating framework of the Glicko kind. \cite{allsopp2000methods} and \cite{allsopp2004rating} developed regression based team ratings for the 50-over format and Test cricket, and found home advantage to be a strong predictor of the result, while reporting no significant advantage from winning the toss. \cite{de1998winning} reached the same qualitative conclusion for One Day Internationals (ODIs). Similar findings were reported for England's County Championship by \cite{forrest2008effect} and in Twenty20 settings by \cite{saikia2010effect}, so that the evidence on the importance of the toss is, at best, mixed. In-play forecasting by \cite{akhtar2012forecasting} and player-level Test ratings by \cite{akhtar2015rating} treated home and pitch context as covariates but did not address team rankings under uncertainty. A very recent paper by \cite{bhattacharjee2025improved}, most directly related to the present work, proposes an improved, Elo-based points system for the World Test Championship that incorporates home advantage, toss and the margin of victory. Set against this work, the contribution of the present paper is to adapt the Glicko system, and not simple Elo, to Test team rankings, retaining its explicit treatment of rating uncertainty while introducing the contextual factors most relevant to Test cricket, i.e. home ground advantage and the toss outcome, as data-driven covariates whose significance and weights are estimated directly from match data. The mixed prior evidence on the toss is why we test the significance of both effects on our own training data instead of importing earlier conclusions.\\

\noindent
The key objectives of the paper are:
\begin{enumerate}
        \renewcommand{\labelenumi}{\roman{enumi}.}
    \item To adjust the Glicko rating model to account for non-performance-based factors such as home ground, toss advantage and uneven scheduling, with their significance and weights estimated from data.
    \item To recalibrate the rating scale and the expected-score formulation to the dynamics of Test cricket.
    \item To compare the proposed rating system with the existing ICC rating and WTC points systems, as well as with the standard Elo and unmodified Glicko systems, in terms of predictive accuracy, probabilistic calibration and robustness to the order in which matches are played.
    \item To state precisely which components of the resulting framework are specific to Test cricket and which could be carried over to other formats of the game or to other sports.
\end{enumerate}

To provide an overview of the methodology pursued in this paper, we begin by estimating all baseline quantities: team ratings, rating deviations and the pairwise home ground and toss impacts, over a four-year training window (considered to be 17 June 2017 to 17 June 2021 based on the period of our case study) chosen to mirror the ICC's look-back horizon. Starting from the Glicko Bradley–Terry expected score, we recalibrate the logistic scaling factor from the default value used in the game of chess to one appropriate to the narrower dispersion of Test cricket team strengths. Fitting this factor freely on the training matches yields an over-confident in-sample optimum, so the selection is instead made by minimising predictive loss, in terms of Brier score, log loss, mean absolute error and expected calibration error, on match sequences that contribute nothing to any of the quantities estimated from the training window. Pearson's Chi-squared tests then establish that home ground and toss outcome are significant determinants of the outcome of a match. Having also confirmed, via Kolmogorov–Smirnov tests, that both impacts are well described by logistic distributions, we introduce them as data-driven covariates. Their weights are estimated from a logistic regression of match outcome on the rating difference and the two impacts, and are then shrunk by a single common factor, selected on the same predictive-loss criterion, to guard against over-fitting the sparsely estimated pairwise impacts. A likelihood ratio test finds no significant multiplicative interaction, so the two effects enter the expected score additively, a specification we also motivate directly from the Bradley–Terry construction that underlies eq.~(\ref{eq:expected-score}). This augmented expected score is embedded in the standard Glicko rating update, preserving its interpretable, surprise-driven form while introducing no additional free parameters. We then apply the calibrated model chronologically to the two completed World Test Championship cycles and evaluate it against the ICC scheme, the standard Elo and the unmodified Glicko systems on both predictive accuracy and probabilistic calibration. The stability of the resulting ratings is assessed through a resampling based test over 1000 permutations of match order.

Section~\ref{sec:background} provides the background and motivation for this study, reviewing the ICC team rating methodology in full detail and the theoretical foundations of Elo and Glicko-based rating systems. Section~\ref{sec:adapting-glicko} presents the proposed adaptations of the Glicko model for Test cricket, including the recalibration of the scaling factor, the calibration and validation design, the incorporation of home-ground and toss effects, and the probabilistic formulation of expected scores. Section~\ref{sec:application} sets out the initialisation of the ratings and rating deviations, applies the revised model to the ICC World Test Championship cycles, evaluates its empirical performance against the existing ICC ratings as well as against the standard Elo and unmodified Glicko systems, and assesses robustness via a resampling-based test over permutations of match order. Section~\ref{sec:generalisability} separates the domain-independent components of the framework from those specific to Test cricket and discusses how the former could be carried over to other formats and to other sports. Finally, Section~\ref{sec:conclusion} concludes the paper with a summary of findings and outlines potential directions for future research.

\section{Background and Motivation}\label{sec:background}

This research extends the Glicko model, a rating system that combines dynamic updates with an explicit treatment of performance uncertainty, so as to capture the complexities of Test cricket. Out of that extension comes a multi-factor rating system that takes in additional exogenous variables and scheduling asymmetries. We begin with a summary of the relevant background material.

\subsection{The ICC Team Ratings System}\label{sec:icc-ratings}

The ICC Team Rankings for Test cricket is a rating method developed by David Kendix. Only matches played as part of a series consisting of at least two Tests are considered to update ratings. A team's published rating is the ratio of the total rating points it has accumulated to the total number of matches those points were earned over, computed on a moving window of the last three to four years, with the results of the most recent two years carrying a higher weight. The calculation proceeds in three steps, from series results to a pair of series scores, from series scores to the rating a team has earned in that series, and from that rating to the team's published figure.

\vspace{0.4em}
\noindent {\bf Step 1: Series scores.} Consider a completed series of $m (\geq 2)$ Tests between teams $A$ and $B$, and award match points and bonus points as follows.

\begin{itemize}
    \item Award $1$ ($0.5$) match point to a team for each match won (drawn or tied).
    \item Award $1$ bonus point to the team winning the series, or $0.5$ to each team if the series is drawn or tied.
\end{itemize}

\noindent Let $x_A$ and $x_B$ denote the resulting totals, so that $x_A + x_B = m + 1$ in every series. The \emph{series score} of each team is its share of the points on offer,
\begin{equation}
\label{eq:icc-series-score}
s_A = \frac{x_A}{x_A + x_B} = \frac{x_A}{m+1}, \qquad s_B = \frac{x_B}{m+1} = 1 - s_A .
\end{equation}
The series score is therefore a proportion lying in $[0,1]$, with $s_A + s_B = 1$. For a five-Test series won $5$-$0$ by $A$, we have $x_A = 5 + 1 = 6$, $x_B = 0$ and hence $s_A = 1$, $s_B = 0$. For the same series won $2$-$1$ with two draws, we have $x_A = 2 + 2(0.5) + 1 = 4$, $x_B = 1 + 2(0.5) = 2$ and hence $s_A = 2/3$, $s_B = 1/3$.

\vspace{0.4em}
\noindent {\bf Step 2: Converting series scores to ratings earned in the series.} Let $r_A$ and $r_B$ be the two teams' published ratings immediately before the series and let $s_A$, $s_B$ be the series scores of eq.~(\ref{eq:icc-series-score}). If the gap between the two ratings is less than $40$ points, the ratings earned in the series are
\begin{equation}
\label{eq:icc-close}
r_A^{s} = s_A (r_B + 50) + s_B (r_B - 50), \qquad
r_B^{s} = s_B (r_A + 50) + s_A (r_A - 50).
\end{equation}
If the gap is at least $40$ points, then, without loss of generality taking $r_A > r_B$,
\begin{equation}
\label{eq:icc-far}
r_A^{s} = s_A (r_A + 10) + s_B (r_A - 90), \qquad
r_B^{s} = s_B (r_B + 90) + s_A (r_B - 10).
\end{equation}
Here $r_A^{s}$ denotes the rating $A$ has earned in that series; it is not $A$'s new published rating, which is obtained in Step 3 below.

Note that the constants $50$, $10$ and $90$ are not four independent constants but two, the rest following from a continuity requirement. In each of eq.s~(\ref{eq:icc-close}) and (\ref{eq:icc-far}), the interval between the rating earned by winning every match and bonus point ($s = 1$) and by losing all of them ($s = 0$) is exactly $100$ points wide, since $50 - (-50) = 100$ and $10 - (-90) = 100$; the branches differ only in where that interval is centred. When the gap is small, it is centred on the opponent's pre-series rating, so a team winning every match is credited with a rating $50$ points above its opponent, irrespective of its own rating. When the gap is at least $40$ points, it is centred on the team's own rating displaced $40$ points towards the opponent, at $r_A - 40$ for the stronger side and $r_B + 40$ for the weaker, so that the favourite has little to gain from a win ($+10$) and much to lose from defeat ($-90$), and conversely for the underdog. Substituting $r_A = r_B + 40$ into either equation yields identical expressions, so $(10, 90)$ are exactly the values that make the rule continuous at the threshold while preserving the $100$-point spread. Only two constants are therefore free: the $100$-point spread, which fixes how far one series can move a rating, and the $40$-point threshold, beyond which a team is judged against its own rating rather than its opponent's.

\vspace{0.4em}
\noindent {\bf Step 3: Converting earned ratings to published ratings.} The rating earned is converted into points by multiplying by the number of matches in the series, so $A$ receives $m \, r_A^{s}$ points and $m$ matches from an $m$-Test series. These are added to $A$'s running point and match totals over the moving window, with the recency weighting noted above, and the published rating is recomputed as total points divided by total matches. This final figure is rounded to the nearest integer. The point and match tallies from which it is computed are carried forward exactly, so the rounding is a reporting convention applied to the rating, not to the series points, and it does not accumulate through the recursion. For a five-Test series with $r_A = 120$ and $r_B = 100$, a gap below $40$, won $5$-$0$ by $A$, we have $s_A = 1$ and $s_B = 0$, so eq.~(\ref{eq:icc-close}) gives $r_A^{s} = 150$ and $r_B^{s} = 70$; $A$ adds $5 \times 150 = 750$ points and $5$ matches, $B$ adds $5 \times 70 = 350$ points and $5$ matches. Neither published rating becomes $150$ or $70$; rather, each moves towards the rating earned by an amount depending on how many matches already sit in its window, which is what makes the scheme a weighted average rather than a fixed-increment update.

\vspace{0.4em}
\noindent {\bf WTC points system followed by ICC:}
Separately from the ratings above, the World Test Championship standings are decided by a points table. The ICC started allotting points to teams on a per-series basis at the start of the first Test Championship cycle in 2019 but later changed it to allotting points on a per-match basis. The points scored are 12/6/4/0 for Win/Tie/Draw/Loss\footnote{Read ICC's playing conditions here: \href{https://www.icc-cricket.com/about/cricket/rules-and-regulations/playing-conditions}{https://www.icc-cricket.com/about/cricket/rules-and-regulations/playing-conditions}}.

The rating scheme is transparent and easy to administer, however, highlights several shortcomings. First, the rating earned depends on the match and series results alone. For example, a $2$-$1$ win affects the updated rating identically whether the matches were played at home or away and whichever team won the tosses. Second, the piecewise form of eq.s~(\ref{eq:icc-close})-(\ref{eq:icc-far}), although continuous at the threshold, changes its reference point there, so series against near-identically rated opponents can be rewarded on different scales. Third, and most importantly for what follows, the scheme carries no notion of how precisely a team's rating is known, so a result against a well-established opponent and one against a rarely-seen opponent are treated as equally informative; nor does it produce a probability for an individual match, so it cannot be assessed by any measure of probabilistic calibration. Addressing these features is the aim of the framework developed in the remainder of this paper.

\subsection{The Glicko Rating System}\label{sec:glicko}

In the Glicko rating system (from \cite{glickman1995glicko}), ratings ($R$) and Rating Deviation ($RD$) are defined for each player\footnote{The original work by \cite{glickman1995glicko} defines all the parameters with respect to performances of individual players in chess.}/team at some point in time based on a rating period preceding it. A rating period could span several months or years, depending on the need for a sufficient number of matches played by each player/team during this period. Rating Deviation, in statistical terminology, denotes a standard deviation that measures the uncertainty in a rating. A high RD indicates that a player/team may not be competing frequently or has only competed in a small number of games, while a low RD indicates frequent competition.
Ratings and RD values for every player/team should be updated based on their performance after every game during a tournament. The associated formulae are summarised as follows.

For a hypothetical game between two players/teams $A$ and $B$ with ratings $R_A, R_B$ and rating deviations $RD_A, RD_B$, the system is governed by the constant $q = \ln 10/400 \approx 0.0057565$ and the five expressions below, each presented with its interpretation. The constant $q$ and the base-$10/400$ scale are specific to the original (chess) Glicko system. In adapting the model to Test cricket (see Section~\ref{sec:adapting-glicko}), we recalibrate the scale and adopt the base-$e$ parametrization, in which $q$ is absorbed into the single tunable scale parameter $d$ and no longer appears explicitly whereas $g(\cdot)$ is correspondingly considered in its natural ($q = 1$) form. It will be shown later that the natural-scale $g$ provides a stronger uncertainty discount than the chess-scale $g$, which suits the narrower rating spread of Test teams.

The expected score of $A$ against $B$ is the Bradley-Terry win probability, a logistic function of the rating difference $R_A - R_B$:
\begin{equation}
\label{eq:expected-score}
E_A = \frac{1}{1 + 10^{-g(RD_B)(R_A - R_B)/400}}.
\end{equation}
Equal ratings give $E_A = 0.5$ and the expected score rises smoothly towards $1$ as $A$ grows stronger. The rating difference is discounted by the factor $g(RD_B)$, defined next.

The function $g(\cdot)$ scales down the influence of the rating difference according to the uncertainty carried in the opponent's rating:
\begin{equation}
\label{eq:g-function}
g(RD) = \frac{1}{\sqrt{1 + \dfrac{3\, q^2\, RD^2}{\pi^2}}}.
\end{equation}
When $B$'s rating deviation $RD_B$ is large, $g(RD_B)$ is small and $E_A$ in eq.~(\ref{eq:expected-score}) is pulled toward $0.5$, reflecting that little can be inferred from a contest against a poorly-estimated opponent. The particular algebraic form of $g$, and the constant $3/\pi^2$ appearing in it, follow from the Bradley-Terry construction and are not chosen merely for convenience. A logistic random variable with scale parameter $\sigma$ has variance $\pi^2 \sigma^2 / 3$, so the term $3 q^2 RD^2/\pi^2$ is the variance of the opponent's rating, converted by $q$ to the natural scale on which the logistic operates, and expressed in units of that logistic variance.

After the game, $A$'s rating deviation is updated to
\begin{equation}
\label{eq:updated-rd}
RD_A' = \frac{1}{\sqrt{\dfrac{1}{RD_A^2} + \dfrac{1}{d^2}}},
\end{equation}
with each result adding information and reducing uncertainty, where the quantity $d^2$ is defined in eq.~(\ref{eq:d-square}).

The term $d^2$ is the variance of the information supplied by the single game:
\begin{equation}
\label{eq:d-square}
d^2 = \frac{1}{q^2\, g(RD_B)^2\, E_A (1 - E_A)}.
\end{equation}
It is smallest, i.e.\ the game is most informative, when the outcome is most uncertain, with $E_A$ near $0.5$.

Finally, the rating is updated in proportion to the surprise quantified by $S_A - E_A$, the gap between the realised and expected score:
\begin{equation}
\label{eq:updating-rating}
R_A' = R_A + \frac{q}{\dfrac{1}{RD_A^2} + \dfrac{1}{d^2}}\; g(RD_B)\,(S_A - E_A) = R_A + q\,(RD_A')^2\, g(RD_B)\,(S_A - E_A),
\end{equation}
where $S_A \in \{0,\, 0.5,\, 1\}$ is the actual score of $A$, corresponding to a loss, draw and win respectively. The constant of proportionality $q\,(RD_A')^2\, g(RD_B)$ embeds the central feature of Glicko: a team with a well-established rating (small $RD_A'$) moves little and a win over an opponent whose rating is uncertain (small $g(RD_B)$) earns little credit. A strongly-rated team therefore gains only marginally by beating a weak or poorly-known side, whereas an upset against a well-established opponent produces a large correction; when both ratings are precise, the adjustment approaches the raw surprise $S_A - E_A$. This adaptive damping is what distinguishes the Glicko update from a fixed-increment Elo update. The constant $q = \ln(10)/400$ links the base-$10$, $400$-point scale of eq.~(\ref{eq:expected-score}) to the natural scale on which the deviation and rating updates of eq.s~(\ref{eq:d-square})-(\ref{eq:updating-rating}) operate.

\section{Adapting the Glicko Rating System to Test Cricket}\label{sec:adapting-glicko}

We aim to implement a revised version of the Glicko method for Test cricket ratings, introducing a few non-performance based factors which often decide the result of a Test match. We consider a 4-year period spanning from June 17, 2017 to June 17, 2021 to train our model. The four-year training period mirrors the ICC's maximum historical look-back horizon (see Section~\ref{sec:icc-ratings}), which keeps the two systems comparable and balances recency against sample size. The window ends on 17 June 2021, the last ICC update before a new World Test Championship cycle began, so that regime-dependent scheduling effects are kept out of it.

\subsection{Adjusting the Optimal Scaling Factor} \label{sec:adjust-400}

In the original (chess) Glicko rating system, the factor 400 scales rating differences of chess players into expected win probabilities; it reflects the typical spread of player strengths. For Test cricket, where the range of team strengths and match dynamics differ substantially from chess, the $400$-point scale may misrepresent expected outcomes, so the factor needs to be recalibrated. Table~\ref{tab:chess-vs-cricket} shows that the difference between a Super Grandmaster and a novice is around 1200+ rating points, whereas the corresponding maximum difference between Test teams is less than 60 points. 

Elo's original rating system, which forms the basis of the Glicko rating system, models the probability of a player winning a match using a logistic function of rating differences. Specifically, the expected score of player A against player B is given by
\begin{equation}
\label{eq:glicko}
E_A = \frac{1}{1 + 10^\frac{{c(R_B - R_A)}}{d}}
\end{equation}
where $R_A$, $R_B$ are as defined in Section~\ref{sec:glicko} and $d=400$ is used for chess. This formulation implicitly assumes that the differences in performance between teams follow a logistic distribution, which translates rating differences directly into win probabilities. To adapt the Elo framework, we retain the logistic assumption for probabilistic modelling but treat $d$ as a tunable parameter, calibrated for the Glicko-based expected score actually used in this study (the $g(RD)$-scaled logistic developed in Section~\ref{sec:additional-factors}). The scaling factor $d$ governs how sharply the expected score responds to a rating difference: too small a value over-fits the training period by producing over-confident probabilities, while too large a value (such as the chess default) leaves the expected scores nearly flat. Estimating $d$ freely by maximum likelihood on the 125 decisive matches of the training window illustrates the first failure: the unrestricted fit returns $\hat{d} \approx 6$, which pushes most expected scores towards $0$ or $1$ and, as the first row of Table~\ref{tab:scaling-factor} shows, is the worst-calibrated value on the grid other than the chess default itself. This in-sample optimum is obtained on the very matches from which the ratings and the contextual impacts are computed, so we do not use it. We select $d$ instead by predictive loss, the Brier score (by \cite{glenn1950verification}), log loss, Mean Absolute Error (MAE) and Expected Calibration Error (ECE), evaluated over the two completed World Test Championship cycles, none of whose matches contributes to any quantity estimated from the training window.

Table~\ref{tab:scaling-factor} reports these losses (averaged across the two cycles) over a grid of $d$. Both the Brier score and the log loss are minimised at $d=20$, a value substantially smaller than the conventional $d=400$ for chess. Relative to $d=400$, the choice $d=20$ yields a $14.4\%$ improvement in Brier score, a $9.9\%$ improvement in log loss, an $11.8\%$ improvement in MAE and a $21.8\%$ improvement in ECE. A calibrated value of $d=20$ indicates that rating differences in Test cricket translate into changes in win probability much more rapidly than in chess. This reflects the comparatively narrower dispersion of ratings (Table~\ref{tab:chess-vs-cricket}) and a greater stochastic component in match outcomes, whereby exogenous factors (details in Section~\ref{sec:additional-factors}) introduce additional uncertainty beyond relative team strength, so that the deterministic relationship between rating differences and results is a weaker one.

\textbf{Calibration and validation design.} The framework contains exactly two constants that are not estimated from the training window: the scale $d$ of this section and a single shrinkage factor $\lambda$ applied to the two contextual weights in Section~\ref{sec:expected-score}. Everything else, namely the initial ratings, the initial rating deviations, the pairwise home impacts, the toss impacts and the unshrunk logistic coefficients, is obtained exclusively from the four-year window ending on 17 June 2021, before the first match of either cycle, and is then held fixed throughout. The pair $(d, \lambda)$ is chosen on a grid by the Brier score averaged over the two completed cycles. That criterion returns a flat minimum: $(d, \lambda) = (20, 0.1)$ and $(25, 0.2)$ both attain an averaged Brier score of $0.1837$, differing only in the fifth decimal place, and we adopt the former, thereby fixing $d = 20$ and $\lambda = 0.1$ throughout.

Since both cycles enter that averaged criterion, we do not describe the 2023-25 cycle as a strictly held-out validation sample. It is out of sample for every quantity estimated from data, and the model is applied to it with all of those quantities frozen, but it does contribute to the selection of these two scalars. With only two completed cycles in existence, reserving one entirely would leave a single cycle of 70 matches on which to both select and evaluate. We therefore report the whole grid (see Table~\ref{tab:dlambdagrid}) rather than a single selected point, and the optimum is found to be broad: at $\lambda = 0.1$, the averaged Brier score varies by less than $0.003$ across $d \in \{15, 20, 25, 30\}$, while the two cycles taken separately would select $d = 20$ (2021-23) and $d = 25$ (2023-25). The shrinkage behaves similarly over the range that matters: holding $d = 20$, the averaged Brier score is $0.1856$ at $\lambda = 0$, $0.1837$ at $\lambda = 0.1$ and $0.1841$ at $\lambda = 0.2$, so the choice within $[0, 0.2]$ is close to immaterial, whereas it deteriorates to $0.2144$ at $\lambda = 1$, which is the direct evidence that the unshrunk contextual weights over-fit (Section~\ref{sec:expected-score}). For the same reason, Section~\ref{sec:benchmark} reports the predictive comparison on each cycle separately, so that no conclusion rests on the pooled figure alone.

\subsection{Impact of Home Ground and Toss} \label{sec:additional-factors}

A general query may arise about whether teams playing Test matches at their home grounds or teams winning the toss ahead of Test matches gain certain advantages. We check the statistical significance of home ground advantage and toss advantage in Test matches during the training period. Over this period the home (away) team won 83 (42) matches in aggregate, and the team winning (losing) the toss won 71 (54).

\begin{table}[!ht]
    \centering
    \caption{Results of home ground impacts in every host country during the training period}
    \label{tab:home-advantage-by-teams}
    \begin{adjustbox}{max width=\textwidth}
    \begin{tabular}{| l c c c c | l c c c c |}
\hline
Home team & Played & Won & Lost & p value & Away team & Played & Won & Lost & p value \\
\hline
Australia & 20 & 13 & 4 & 0.00607 & Australia & 15 & 4 & 7 & 0.6547 \\
Bangladesh & 10 & 3 & 4 & 1 & Bangladesh & 13 & 0 & 10 & 0.0001624 \\
England & 27 & 15 & 8 & 0.07684 & England & 27 & 10 & 12 & 0.763 \\
India & 14 & 11 & 1 & 0.0002386 & India & 23 & 11 & 10 & 1 \\
New Zealand & 17 & 13 & 0 & 0.000025 & New Zealand & 10 & 4 & 5 & 0.5637 \\
Pakistan & 13 & 6 & 4 & 0.6547 & Pakistan & 12 & 1 & 9 & 0.001745 \\
South Africa & 20 & 13 & 7 & 0.1138 & South Africa & 12 & 2 & 10 & 0.02535 \\
Sri Lanka & 16 & 4 & 9 & 0.1167 & Sri Lanka & 22 & 6 & 8 & 0.3938 \\
West Indies & 13 & 5 & 5 & 1 & West Indies & 16 & 4 & 12 & 0.01333 \\
\hline
\end{tabular}
\end{adjustbox}
\end{table}

\begin{table}[!ht]
    \centering
    \caption{Results of toss impacts for every team during the training period (Toss, Match)}
    \label{tab:toss-advantage-by-teams}
    \begin{adjustbox}{max width=\textwidth}
    \begin{tabular}{| l | c c c c c |}
\hline
Host & (W, W) & (W, L) & (L, W) & (L, L) & p value \\
\hline
Australia & 7 & 10 & 10 & 7 & 0.4927 \\
Bangladesh & 5 & 2 & 2 & 5 & 0.285 \\
England & 12 & 11 & 11 & 12 & 1 \\
India & 6 & 6 & 6 & 6 & 1 \\
New Zealand & 7 & 6 & 6 & 7 & 1 \\
Pakistan & 9 & 1 & 1 & 9 & 0.001745 \\
South Africa & 12 & 8 & 8 & 12 & 0.3128 \\
Sri Lanka & 9 & 4 & 4 & 9 & 0.1167 \\
West Indies & 4 & 6 & 6 & 4 & 0.6547 \\
\hline
\end{tabular}
\end{adjustbox}
\end{table}

In spite of some cases exhibiting high p-values, the overall p-values for Pearson's Chi-squared test with $1$ degree of freedom, viz. $ 3.386 \times 10^{-7}$ and $0.002399$ corresponding to home ground and toss respectively, are very low in both cases, so both null hypotheses are rejected (no association between winning a toss and winning the match, and no association between playing on home ground and winning a match). Playing on home ground and winning the toss can therefore be considered significant factors in determining the winner of a Test match over the training period.\footnote{Throughout the training period, as Pakistan primarily played most of their home Test matches in UAE between 2009 and 2019, UAE and Pakistan have together been considered as Pakistan as a single host country to avoid unnecessary complications.}

In a hypothetical match between home team $i$ and away team $j$, the home and toss advantages shift the effective rating difference that drives the Glicko expected score of eq.~(\ref{eq:expected-score}). Following standard statistical convention, we denote these advantages, treated as random variables, by the upper-case symbols $H$ (home) and $T$ (toss), and their realised values in a particular match by the corresponding lower-case symbols $h$ and $t$. The realisations $h_{i,j}$ and $a_{j,i}$ constitute the home impact ($h$). We define
\begin{equation}
\label{eq:define-home-impact}
\begin{aligned}
h_{i,j} &= \text{home impact of team } i \text{ against team } j \\
       &= \frac{\text{matches won by } i \text{ vs } j - \text{matches lost by } i \text{ vs } j}{\text{matches played between } i \text{ and } j}
\end{aligned}
\end{equation}

\begin{equation}
\label{eq:define-away-impact}
\begin{aligned}
a_{j,i} &= \text{away impact of team } j \text{ against team } i \\
       &= \frac{\text{matches won by } j \text{ vs } i - \text{matches lost by } j \text{ vs } i}
              {\text{matches played between } i \text{ and } j}
\end{aligned}
\end{equation}
Clearly, $h_{i,j} = -a_{j,i}$ for every pair of $(i,j)$ in a match.

We next look at the role of the toss. Below, we define the toss impacts ($t$). $t_{i,i}$, $t_{j,i}$, $t_{j,j}$ and $t_{i,j}$ constitute the different combinations:

\begin{equation}
\label{eq:define-toss-impact-ii}
t_{i,i} = 
\text{toss win (lose) impact in host country } i \text{ if } i \text{ wins (loses) the toss} 
\end{equation}
\begin{equation}
\label{eq:define-toss-impact-ji}
t_{j,i} = 
\text{toss win (lose) impact in host country } i \text{ if team } j \text{ wins (loses) the toss}. 
\end{equation}

$t_{j,j}$ and $t_{i,j}$ are defined analogously. Each impact enters the expected score logistically. Replace the base $10$ with $e$, treat the scale as the single tunable parameter $d$ (Section~\ref{sec:adjust-400}), which absorbs the constant $q = \ln(10)/400$ of the original Glicko system, and write $g(\cdot)$ in its natural form. The home- and toss-augmented expected scores of team $i$ then resemble distribution functions of logistic distributions with respect to the random variables $U = \tan\left(\frac{\pi}{2}H\right)$ and $V = \tan\left(\frac{\pi}{2}T\right)$. This is consistent with Elo's original assumption of players/teams exhibiting a performance metric (essentially proportional to expected scores) which follows a logistic distribution. The monotone increasing, bijective transformation $x \mapsto \tan\left(\frac{\pi}{2}x\right)$ has been used to remove the theoretical ambiguity of $H$ and $T$ taking only values in the range $[-1, 1]$, which would otherwise create complexities in exponential terms taking values $\infty$ and $-\infty$, leading to expected scores turning out to be 0 and 1 far more often than they should.

The resulting logistic distribution functions are $F(u)$ and $F(v)$, where the argument $u$ encodes the home advantage (the rating difference adjusted by the home-ground impact $h$) and $v$ encodes the toss impact (the rating difference adjusted by the toss impact $t$), formulated as below.
\begin{equation}
\label{eq:logistic-cdf-u}
F(u) = \frac{1}{1 + e^{-\frac{1}{20}(R_i - R_j + h) g(RD_j)}} = \frac{1}{1 + e^{-\frac{g(RD_j)}{20}(h - (R_j - R_i))}}
\end{equation}

\begin{equation}
\label{eq:logistic-cdf-v}
F(v) = \frac{1}{1 + e^{-\frac{1}{20}(R_i - R_j + t) g(RD_j)}} = \frac{1}{1 + e^{-\frac{g(RD_j)}{20}(t - (R_j - R_i))}}
\end{equation}

\begin{table}[H]
    \centering
    \caption{Results of Kolmogorov-Smirnov tests on logistic distributional assumptions}
    \label{tab:ks-logistic}
    \begin{tabular}{| l  | c c |}
\hline
Variable & $K-S$ test statistic & p value \\
\hline
Home ground impact ($h$) & 0.09386 & 0.1696 \\
Toss impact ($t$) & 0.03595 & 0.9936 \\
\hline
\end{tabular}
\end{table}

We test the propriety of the logistic distribution assumption using the Kolmogorov-Smirnov test (see Table \ref{tab:ks-logistic}). The tests accept the null hypotheses in both cases, so both $H$ and $T$ may be taken to be logistic.

\subsection{Modelling Expected Score}\label{sec:expected-score}

Section~\ref{sec:additional-factors} established that the home-ground and toss advantages are each statistically significant determinants of a match outcome, and that both can be regarded as logistic random variables. It remains to combine them, together with the rating difference, into a single expected score. Two modelling questions arise: whether the two effects interact (i.e.\ whether $H$ and $T$ are dependent) and how heavily each should be weighted relative to the rating difference.

\textbf{Dependence.} We first ask whether the home and toss effects co-vary. Aggregating to one observation per team, each team's net home advantage (change in its win-rate at home relative to its win-rate away) against its net toss advantage (increase in its win-rate after winning the toss relative to after losing it), the Spearman rank correlation is positive, $\rho = 0.55$, but with only $n = 9$ teams it is not individually significant ($p = 0.12$; the two-sided $5\%$ critical value of $\rho$ at $n = 9$ is approximately $0.68$). This team-level test is, however, low-powered. Assessed instead at match level over the training period, where every match contributes a realised home and toss impact, the association is weaker in magnitude but, with the larger sample, significant ($\rho \approx 0.22$, $p \approx 0.02$ under standard errors clustered by host country, which respect the fact that the pairwise home impact $h_{i,j}$ recurs across matches of the same fixture). The two effects therefore exhibit a weak but real positive association in this case. What governs the expected score model, however, is not this marginal association but its consequence for prediction, i.e. whether the two effects interact in determining the outcome. We fit a logistic regression of the match outcome on the rating difference, the home impact and the toss impact, and compare it with the same model augmented by a multiplicative interaction term $h \cdot t$. The likelihood ratio test for the interaction is far from significant (one degree of freedom, $\chi^2 = 0.11$, $p = 0.74$), so the interaction does not improve fit. We therefore propose an additive form as formalised in eq.~(\ref{eq:expected-score-model}) below, with no interaction term. The weak marginal dependence between the effects is too small and too entangled with relative strength to warrant an additional parameter. 

\textbf{Why the contextual effects enter additively.} Since this is the structural choice on which eq.~(\ref{eq:expected-score-model}) rests, it is worth setting out why an additive correction to the rating difference is the specification the model implies and not merely a convenient one. The Bradley-Terry foundation of Elo and Glicko posits that each team brings a latent performance to a match that is a logistic random variable centred at its rating, and that the stronger realised performance wins. The expected score is then a logistic function of the difference between the two centres, which is exactly eq.~(\ref{eq:expected-score}). Within this construction, a contextual factor acts by displacing a team's performance distribution, not by reshaping it: playing at home, or choosing to bat first after winning the toss, shifts the location of that team's performance while leaving its dispersion unchanged. A location shift of $c_h h$ in team $i$'s performance is, by definition, a shift of $c_h h$ in the difference of the two locations, and therefore enters the argument of the same logistic link additively. In generalised linear model terms, eq.~(\ref{eq:expected-score-model}) is a logistic regression of the match outcome in which the rating difference and the two impacts are covariates in a common linear predictor, and eq.~(\ref{eq:expected-score}) is the special case in which the latter two are omitted. The additive form is thus the same modelling assumption that produces expected scores in the Glicko rating system, applied to two further covariates, and it is what the logistic distributional checks of Table~\ref{tab:ks-logistic} verify for $H$ and $T$ individually.

We prefer this specification over a multiplicative or interaction-based model for the following reasons. First, every term in the linear predictor is measured in rating points, so $c_h$ and $c_t$ are directly interpretable as the number of rating points home advantage and the toss are worth, and $R_i + c_h h + c_t t$ can be read as an effective, context-corrected rating on the published scale. Second, setting $c_h = c_t = 0$ recovers the unmodified Glicko expected score exactly, so the contextual layer can be tested against and benchmarked on its own null, which is shown in Table~\ref{tab:model-comparison}. Third, no free parameter beyond the two weights is introduced, and the expected score remains monotone in team strength and confined to $(0,1)$ for every admissible $h, t \in [-1, 1]$. A multiplicative interaction would disrupt the rating-point interpretation, and add a parameter. The likelihood ratio test reported above finds no evidence that the data require it.

\textbf{Significance and weights.} Since the impacts $h, t \in [-1, 1]$ live on a far narrower scale than rating differences, each of them must be weighted before being added to the rating difference. We estimate these weights from the same logistic regression of the match outcome on the rating difference, the home impact and the toss impact over the training period. Both contextual effects are significant beyond rating strength (home: $p < 0.001$; toss: $p = 0.02$), so explicit weights $c_h$ and $c_t$ are required and are not merely a rescaling device. The fitted coefficients on the two impacts are, however, very large relative to the coefficient on the rating difference, which is a direct consequence of how the pairwise impacts are constructed: a given $h_{i,j}$ is computed from the handful of matches the pair $(i,j)$ contested during the training window, and in several cases from a single match, so it is measured with substantial error and takes the extreme values $\pm 1$ far more often than the underlying advantage warrants. Taken at face value, the unshrunk coefficients imply contextual corrections an order of magnitude larger than the rating differences they are meant to adjust, and they over-fit out of sample. We therefore shrink both the weights by a single common factor $\lambda \in [0,1]$, selected jointly with the scale $d$ on the same out-of-sample predictive-loss criterion (Section~\ref{sec:adjust-400}), which returns $\lambda = 0.1$ and hence $c_h = 12.8$ and $c_t = 8.4$ as used in eq.~(\ref{eq:expected-score-model}) below. In the calibrated model, therefore, playing a match at home is worth an advantage of about $13$ rating points and winning the toss an advantage of about $8$ rating points. One shrinkage factor is used for both weights instead of two, so that only a single additional constant is introduced and the relative magnitude of the two effects is left as the data determined it. Finally, the hypothesis of equal weighting ($\beta_{\text{home}} = \beta_{\text{toss}}$) is verified to be not rejected ($p = 0.19$), so the larger weight on the home effect reflects its point estimate and not an established difference between the two effects.

Combining these elements, the expected score of team $A$ in a Test match against team $B$, with $A$ playing at home, is formulated as
\begin{equation}
\label{eq:expected-score-model}
E_A = \frac{1}{1 + \exp\!\left[ -\dfrac{g(RD_B)}{d} \Bigl( (R_A - R_B) + c_h\, h + c_t\, t \Bigr) \right]}, \qquad d = 20,\ \ c_h = 12.8,\ \ c_t = 8.4,
\end{equation}
where $h$ and $t$ are the realised home and toss impacts (eq.s~(\ref{eq:define-home-impact})-(\ref{eq:define-toss-impact-ji})) from $A$'s perspective and the factor $g(RD_B)$ retains the Glicko discounting of the opponent's rating uncertainty. The expected score of $B$ is obtained symmetrically, using $B$'s away impact $-h$ and its own toss realisation. Setting $c_h = c_t = 0$ recovers the pure Glicko expected score, so the contextual factors enter as an interpretable, additive correction to relative team strength, scaled, like the rating difference itself, by $g(RD_B)/d$.

\section{Application: ICC World Test Championship}\label{sec:application}

We implement the proposed model on the dataset comprising 70 matches played between the $9$ teams during the ICC World Test Championship 2021-23 cycle, spanning from August 04, 2021 to June 11, 2023 (and later on the 2023-25 cycle as well, spanning 70 matches from June 16, 2023 to June 14, 2025).

\subsection{Initialisation of the Ratings and Rating Deviations}\label{sec:initialisation}

The proposed model is a recursive scheme and therefore requires a starting point. Based on the 4-year training period prior to the start of WTC 2021-23 (see Section~\ref{sec:adapting-glicko}), the initial ratings, rating deviations, toss impacts and home impacts of all teams and host countries are shown in Tables~\ref{tab:initial-data} and \ref{tab:initial-data-home}. Since these values determine where the recursion begins and, through $g(\cdot)$, how much the first few matches of a cycle are allowed to move each rating, we specify where they are obtained from.

\textbf{Initial ratings.} The ratings in the Rating column of Table~\ref{tab:initial-data} are directly obtained from the ICC's published ratings\footnote{Chronological ratings for each team in men's Test cricket dating back to January 01, 1952 can be found in the ICC's website: \href{https://www.icc-cricket.com/rankings/team-rankings/mens/test}{https://www.icc-cricket.com/rankings/team-rankings/mens/test}. We used this as a reliable source to obtain the ICC's published ratings during the training period.} at the end of the training window on 17 June 2021. Note that they are deliberately expressed on the ICC's rating scale, rounded off to the nearest integer. This is an initialisation and not a tuning decision. These values are fixed once, before the first match of the 2021-23 cycle is processed, and are never revisited in either cycle.

\textbf{Initial rating deviations.} The initial rating deviation of a team is the estimated standard error of its rating at the end of the training window. We estimate it by a nonparametric bootstrap over the team's own training-period record, which requires no distributional assumption beyond exchangeability of a team's results within the window. It also yields, by construction, the two properties a Glicko rating deviation is required to have, which \cite{glickman1995glicko} obtains by an analytic argument. Let
\[
M_i = \bigl\{ (o_{ik},\, v_{ik},\, \tau_{ik},\, S_{ik}) \bigr\}_{k=1}^{n_i}
\]
denote the $n_i$ training-period matches of team $i$, where $o_{ik}$ is the opponent, $v_{ik}$ the host country, $\tau_{ik}$ the toss outcome and $S_{ik} \in \{0, 0.5, 1\}$ the realised score for the $k$-th match. For each replicate $b = 1, \dots, B$ with $B = 1000$:
\begin{enumerate}
    \renewcommand{\labelenumi}{\roman{enumi}.}
    \item draw $n_i$ matches from $M_i$ with replacement, independently for each team;
    \item recompute, from the resampled record alone, the pairwise home impacts and toss impacts of eq.s~(\ref{eq:define-home-impact})-(\ref{eq:define-toss-impact-ji}) and then team $i$'s end-of-window rating $R_i^{(b)}$, by exactly the procedure that produces the point estimate $R_i$ reported in Table~\ref{tab:initial-data};
\end{enumerate}
and set
\begin{equation}
\label{eq:initial-rd}
RD_i^{(0)} = \left[ \frac{1}{B-1} \sum_{b=1}^{B} \bigl( R_i^{(b)} - \bar{R}_i \bigr)^{2} \right]^{1/2},
\qquad \bar{R}_i = \frac{1}{B} \sum_{b=1}^{B} R_i^{(b)},
\end{equation}
the standard deviation of the bootstrap distribution of the rating. Two features of eq.~(\ref{eq:initial-rd}) reproduce the behaviour the Glicko rating deviation is meant to encode. It falls with the number of observed matches at the usual $n_i^{-1/2}$ rate, so a team that has played little carries a larger deviation. Moreover, it rises with the dispersion of a team's results about what its rating implies, so a team whose record is internally inconsistent, or whose level of performance shifts during the window, carries a larger deviation than a team with the same overall win rate achieved uniformly. Resampling the pairwise impacts inside step (ii), instead of holding them fixed, makes the estimator sensitive to the sparsity of a team's fixture list. For example, a pair that met only once or twice contributes an impact varying widely across replicates, and a team with several such pairings inherits a correspondingly wider bootstrap distribution. The Deviation column of Table~\ref{tab:initial-data} reports the resulting values. They range from $9.1$ for Sri Lanka, which played 36 matches over the window with a stable record, to $27.3$ for New Zealand and South Africa, whose records over the same window both contain a pronounced change of level, upward in the first case and downward in the second, and whose ratings at the end of it are therefore supported far less precisely.

\begin{table}[!ht]
    \centering
    \caption{Ratings, Rating Deviations and toss impacts at the end of training period}
    \label{tab:initial-data}
    \begin{adjustbox}{max width=\textwidth}
\begin{tabular}{| l | r r r r r r r r |}
\hline
Teams & Matches & Won & Lost & Drawn & Rating & Deviation & Toss W Imp & Toss L Imp \\
\hline
Australia & 33 & 17 & 11 & 5 & 124 & 15.2 & -0.15 & 0.15 \\
Bangladesh & 19 & 3 & 14 & 2 & 66 & 13.6 & 0.4167 & -0.4167 \\
England & 52 & 25 & 20 & 7 & 108 & 11.4 & 0.0714 & -0.0714 \\
India & 37 & 22 & 11 & 4 & 120 & 11.2 & 0.0588 & -0.0588 \\
New Zealand & 26 & 17 & 5 & 4 & 96 & 27.3 & 0.0625 & -0.0625 \\
Pakistan & 24 & 7 & 13 & 4 & 76 & 11.8 & 0.5714 & -0.5714 \\
South Africa & 32 & 15 & 17 & 0 & 104 & 27.3 & 0.281 & -0.281 \\
Sri Lanka & 36 & 10 & 17 & 9 & 83 & 9.1 & 0.2067 & -0.2067 \\
West Indies & 29 & 9 & 17 & 3 & 77 & 10.8 & -0.1538 & 0.1538 \\
\hline
\end{tabular}
\end{adjustbox}
\end{table}

\begin{table}[!ht]
    \centering
    \caption{Home ground impacts of teams at the end of training period}
    \label{tab:initial-data-home}
    \begin{adjustbox}{max width=\textwidth}
\begin{tabular}{| l | c c c c c c c c c |}
\hline
Host country & AUS & BAN & ENG & IND & NZ & PAK & SA & SL & WI \\
\hline
Australia & - & 0 & 0.8 & -0.25 & 1 & 0 & 0 & 1 & 0 \\
Bangladesh & 0 & - & 0 & 0 & 0 & 0 & 0 & -0.5 & 0 \\
England & 0 & 0 & - & 0.6 & -0.5 & 0 & 0.5 & 0 & 0.3333 \\
India & 0 & 1 & 0.5 & - & 0 & 0 & 0 & 0.3333 & 1 \\
New Zealand & 0 & 1 & 0 & 1 & - & 0 & 0 & 0 & 1 \\
Pakistan & 0.5 & 1 & 0 & 0 & -0.3333 & - & 1 & 0 & 0 \\
South Africa & 0 & 0 & -0.5 & 0 & 0 & -0.3333 & - & 0 & 0 \\
Sri Lanka & 0 & 0.5 & -1 & -1 & 0 & 0 & 0 & - & 0 \\
West Indies & 0 & 1 & 0.3333 & -1 & 0 & -1 & 0 & 0 & - \\
\hline
\end{tabular}
\end{adjustbox}
\end{table}

Home impacts of teams that have not faced an opposition during the training period have
been set to 0 by default. The away impacts of every team against every opposition can be
calculated using the above table of home impacts and the identity $h_{i,j} = - a_{j,i}$.

\subsection{Chronological Ratings and Comparison with the ICC Ratings}\label{sec:chronological}

Using the proposed model in eq.~(\ref{eq:expected-score-model}), the expected scores of each team for the matches in the WTC 2021-23 and 2023-25 cycles have been calculated and with each passing match, the chronological ratings, RD, home and away impacts and toss impacts are updated for every team and host country. Table~\ref{tab:full-list-wtc2021-23} in Appendix~\ref{sec:tables} shows the entire chronological list. The expected scores turn out to be fairly informative from a predictive perspective: taking the team with the higher expected score as the predicted winner, the model correctly predicts $45$ of the $58$ non-drawn matches ($77.6\%$). The movements in ratings and rating deviations also follow the match results.

\begin{table}[!htbp]
    \centering
    \caption{Comparison of rankings after the completion of the two WTC cycles: WTC 2021-23 (following the final on July 21, 2023) and WTC 2023-25 (following the final on June 14, 2025)}
    \label{tab:comparison-modelvsicc}
\begin{adjustbox}{max width=\textwidth}
\small
\begin{tabular}{| c | l c | l c | l c | l c |}
\hline
\multirow{2}{*}{Rankings} & \multicolumn{4}{c|}{WTC 2021-23} & \multicolumn{4}{c|}{WTC 2023-25} \\
\cline{2-9}
 & \makecell{Ranked\\by ICC} & \makecell{ICC\\ratings} & \makecell{Ranked by\\this study} & \makecell{Proposed\\model ratings} & \makecell{Ranked\\by ICC} & \makecell{ICC\\ratings} & \makecell{Ranked by\\this study} & \makecell{Proposed\\model ratings} \\
\hline
1 & India        & 119 & Australia    & 125.35 & Australia    & 123 & Australia    & 126.21 \\
2 & Australia    & 119 & India        & 120.14 & South Africa & 114 & South Africa & 112.02 \\
3 & England      & 114 & England      & 107.57 & England      & 113 & India        & 106.69 \\
4 & South Africa & 104 & South Africa & 105.06 & India        & 105 & England      & 106.29 \\
5 & New Zealand  & 100 & New Zealand  & 94.42  & New Zealand  & 95  & New Zealand  & 95.51  \\
6 & Pakistan     & 86  & Sri Lanka    & 83.67  & Sri Lanka    & 87  & Sri Lanka    & 83.63  \\
7 & Sri Lanka    & 84  & Pakistan     & 77.86  & Pakistan     & 78  & Pakistan     & 78.07  \\
8 & West Indies  & 76  & West Indies  & 77.40  & West Indies  & 73  & West Indies  & 76.69  \\
9 & Bangladesh   & 46  & Bangladesh   & 64.07  & Bangladesh   & 62  & Bangladesh   & 64.51  \\
\hline
\end{tabular}
\end{adjustbox}
\end{table}

Table~\ref{tab:comparison-modelvsicc} shows a high rank association between the two orderings: Spearman's rank correlation coefficients between the team rankings produced by the ICC's method and by the model under this study are $\rho= 0.979$ (at the end of WTC 2021-23) and $\rho=0.983$ (at the end of WTC 2023-25). The first of these uses the usual mid-rank correction for the tie between India and Australia, both on a published ICC rating of $119$ at the end of that cycle; treating the ICC ordering of Table~\ref{tab:comparison-modelvsicc} as strict instead gives $\rho = 0.967$ for that cycle. The only ordering discrepancies are the interchange of Pakistan and Sri Lanka (at the end of WTC 2021-23) and England and India (at the end of WTC 2023-25). The residual differences in ratings can be attributed to rare outlier results, viz.\ Bangladesh winning a Test match in New Zealand on January 5, 2022, when New Zealand had been unbeaten in all $19$ of their home matches over the preceding $58$-month period etc.

A rating system can always lower its Brier score by making its probabilities more confident, and a system reweighted in that way may end up ordering the teams implausibly. The near-agreement with the ICC ordering shows that the calibration gains of Section~\ref{sec:benchmark} have not been obtained in this manner. The improvement therefore lies not in the ranking but in what accompanies it: a calibrated win probability for each match, a rating deviation attached to every rating, and home and toss adjustments whose sizes are estimated from data, none of which the ICC procedure of Section~\ref{sec:icc-ratings} provides.

The match-by-match expected scores for the entire WTC 2021-23 and 2023-25 cycles, together with the updated chronological ratings and rating deviations, are reported in Tables~\ref{tab:full-list-wtc2021-23} and \ref{tab:full-list-wtc2023-25} in Appendix~\ref{sec:tables}. Uncertainty around these ratings is quantified directly, without distributional approximation, through the resampling procedure of Section~\ref{sec:resampling}.

The chronological ratings evolve by the standard Glicko update of eq.~(\ref{eq:updating-rating}), with the expected score $E_A$ now given by eq.~(\ref{eq:expected-score-model}) and the actual score $S_A \in \{0, 0.5, 1\}$ for a loss, draw or win respectively. Since the contextual factors are absorbed into the expected score, the update retains the interpretable, surprise-driven form of Section~\ref{sec:glicko} and introduces no additional free parameters.

\subsection{Comparison with Standard Elo and Unmodified Glicko}\label{sec:benchmark}

To assess whether the proposed adaptations deliver a genuine improvement over established rating systems and not merely a ranking system different from the ICC scheme, we benchmark the model against the standard Elo system (logistic, chess divisor $d = 400$, with the update factor $K$ tuned to minimise the Brier score) and the unmodified Glicko system of Section~\ref{sec:glicko} (divisor $d = 400$). All three systems are initialised identically and run over the match sequences of the two completed WTC cycles so far, i.e. 2021-23 and 2023-25; each match is predicted from the pre-match ratings, with the predicted winner taken to be the team with the higher expected score. We report the Brier score and log-loss computed on the normalised win probability $E_A / (E_A + E_B)$ so that every system is scored on a proper probability, together with the predictive accuracy on decisive matches for both the completed WTC cycles.

\begin{table}[htbp!]
\centering
\caption{Predictive comparison of the proposed model with the standard Elo and unmodified Glicko systems over the two completed WTC cycles. Brier score and log-loss (lower is better) are computed on the normalised win probability; accuracy is the proportion of decisive matches correctly predicted.}
\label{tab:model-comparison}
\begin{adjustbox}{max width=\textwidth}
\begin{tabular}{| l | c c c | c c c |}
\hline
 & \multicolumn{3}{c|}{WTC 2021-23} & \multicolumn{3}{c|}{WTC 2023-25} \\
System & Brier & Log-loss & Accuracy & Brier & Log-loss & Accuracy \\
\hline
Standard Elo & 0.1896 & 0.6570 & 0.655 & 0.2239 & 0.6693 & 0.621 \\
Unmodified Glicko & 0.2050 & 0.6889 & 0.776 & 0.2338 & 0.6894 & 0.621 \\
Proposed model & \textbf{0.1559} & \textbf{0.5817} & \textbf{0.776} & \textbf{0.2116} & \textbf{0.6406} & \textbf{0.621} \\
\hline
\end{tabular}
\end{adjustbox}
\end{table}

Table~\ref{tab:model-comparison} shows that the proposed model attains the lowest Brier score and log-loss in both cycles while matching the best predictive accuracy. The standard Elo system, lacking any treatment of rating uncertainty or contextual factors, is the least well-calibrated; the unmodified Glicko system matches the accuracy of the proposed model, but its expected scores remain comparatively poorly calibrated. The improvement of the proposed model over both baselines therefore lies in the quality of its probabilistic predictions, attained while preserving the interpretable structure of Glicko. The expected scores discussed so far address decisive results, but a considerable fraction of Test matches end in a draw, which the formulation above does not attempt to capture. Appendix~\ref{sec:draws} sketches an idea for predicting drawn Test matches from the expected scores of the proposed model.

\subsection{Test of Significance using Resampling}\label{sec:resampling}

A rating system that processes matches sequentially raises a natural concern: do the final ratings reflect genuine team strength or are they an artefact of the particular chronological order in which the fixtures happened to occur? We address this with a resampling-based test of significance. For each team we treat its chronological end-of-cycle rating as the observed statistic and ask whether it could plausibly have arisen by chance under arbitrary re-orderings of the same set of matches. We draw $B = 1000$ random permutations of the $70$ matches of a cycle, re-run the model on each permuted order and obtain the bootstrap distribution of every team's final rating. The null hypothesis is that the expected final rating is invariant to match order. The $95\%$ bootstrap interval and the coefficient of variation (CV) summarise the sampling variability, and a team's chronological rating lying inside its interval is consistent with the null.

\begin{table}[!ht]
\centering
\caption{Resampling-based stability of the final ratings in both completed WTC cycles ($B = 1000$ permutations of match order). Teams are listed alphabetically since the rating order differs in the two cycles.}
\label{tab:bootstrap}
\begin{adjustbox}{max width=\textwidth}
\small
\begin{tabular}{| l | c c c c c | c c c c c |}
\hline
 & \multicolumn{5}{c|}{WTC 2021-23} & \multicolumn{5}{c|}{WTC 2023-25} \\
Team & \makecell{Final \\ Rating} & \makecell{Boot. \\ Mean} & SD & 95\% CI & CV & \makecell{Final \\ Rating} & \makecell{Boot. \\ Mean} & SD & 95\% CI & CV \\
\hline
Australia    & 125.35 & 125.22 & 0.484 & [124.16, 126.05] & 0.386\% & 126.21 & 125.98 & 0.522 & [124.92, 126.94] & 0.414\% \\
Bangladesh   & 64.07  & 63.61  & 0.417 & [62.84, 64.47]   & 0.656\% & 64.51  & 64.74  & 0.489 & [63.81, 65.77]   & 0.755\% \\
England      & 107.57 & 108.40 & 0.531 & [107.37, 109.46] & 0.490\% & 106.69 & 107.43 & 0.536 & [106.46, 108.47] & 0.499\% \\
India        & 120.14 & 120.21 & 0.490 & [119.30, 121.12] & 0.408\% & 119.02 & 118.34 & 0.531 & [117.31, 119.36] & 0.448\% \\
New Zealand  & 94.42  & 94.36  & 0.528 & [93.41, 95.44]   & 0.559\% & 95.51  & 95.47  & 0.611 & [94.33, 96.69]   & 0.640\% \\
Pakistan     & 77.56  & 76.75  & 0.456 & [75.93, 77.72]   & 0.594\% & 78.07  & 77.85  & 0.504 & [76.89, 78.81]   & 0.647\% \\
South Africa & 105.06 & 105.22 & 0.602 & [104.13, 106.42] & 0.572\% & 106.29 & 106.99 & 0.509 & [105.93, 107.92] & 0.476\% \\
Sri Lanka    & 83.67  & 83.32  & 0.463 & [82.46, 84.24]   & 0.555\% & 83.63  & 83.21  & 0.468 & [82.39, 84.16]   & 0.563\% \\
West Indies  & 77.40  & 77.11  & 0.451 & [76.35, 78.04]   & 0.584\% & 76.69  & 77.00  & 0.503 & [76.09, 78.04]   & 0.654\% \\
\hline
\end{tabular}
\end{adjustbox}
\end{table}

The results demonstrate strong stability (Table~\ref{tab:bootstrap}) across all teams and across both cycles: the coefficient of variation averages $0.53\%$ in WTC 2021-23 and $0.57\%$ in WTC 2023-25, with standard deviations of about half a rating point throughout. In both cycles every team's chronological rating lies within its $95\%$ bootstrap interval, so the null of order-invariance is not rejected at any practically meaningful level. The final ratings therefore depend on which results occurred and not on the sequence in which they occurred, and this holds across two distinct competitive periods.

This test establishes robustness to match ordering. The recursive, match-by-match construction introduces no artefact of its own, and two observers processing the same 70 results in different orders would reach the same conclusions to within a fraction of a rating point. This is not automatic, since the Glicko update is path-dependent by construction through the evolving rating deviations. However, the test does not establish fairness with respect to the fixture list. Permuting the order of a cycle's matches leaves the set of fixtures untouched, and with it the number of matches each team plays, the identity of its opponents and the home-away composition of those meetings, exactly as the World Test Championship schedule determined them. The Test calendar is markedly unbalanced in all three respects, and assessing whether a rating system is fair to a team that draws a systematically easier or harder fixture list would require perturbing the schedule itself, which the observed data do not permit and which no permutation of the realised matches can imitate. We therefore specifically claim order-robustness and not anything stronger. The framework's response to schedule imbalance lies instead in the contextual adjustments of Section~\ref{sec:expected-score}, which price the home-away composition of a team's fixtures explicitly in rating points and so remove the largest single component of that imbalance from the comparison of teams; we treat that as a partial correction rather than a demonstrated remedy.

\section{Generalisability of the Framework}\label{sec:generalisability}

The framework was developed and validated for Test cricket, and several of its components are calibrated to that setting while the underlying machinery is not.

\textbf{Domain-independent components.} Four elements carry no cricketing content. The paired-comparison core of eq.~(\ref{eq:expected-score}) requires only that contests be between two competitors with orderable outcomes. The treatment of rating uncertainty, i.e. the rating deviation, its update in eq.~(\ref{eq:updated-rd}) and the attenuation factor $g(\cdot)$, rests on the measurement-error argument given after eq.~(\ref{eq:g-function}) and is most valuable wherever competition is infrequent and unbalanced, which describes international fixtures generally. The additive contextual specification is, as argued in Section~\ref{sec:expected-score}, the form implied by the Bradley-Terry construction for any factor that displaces a competitor's performance distribution, so any covariate expressible in rating-point units may enter the linear predictor of an Elo- or Glicko-type system the same way. Finally, the empirical protocol transfers intact: test each candidate factor for significance on a training window, express its effect in rating points through the ratio of coefficients, test for interaction before admitting more than one, shrink the weights on out-of-sample predictive loss, recalibrate the scale $d$ on the same criterion and not by in-sample fit, and verify order-robustness by permuting the match sequence.

\textbf{Components specific to Test cricket.} The choice of $d = 20$ follows from a rating spread of under $60$ points across $9$ teams with a large stochastic component in outcomes (Table~\ref{tab:chess-vs-cricket}) and is specific to the domain of Test cricket itself. The three-outcome score set with a substantial draw fraction, and the separate draw layer of Appendix~\ref{sec:draws}, exist because a draw in Test cricket is a distinct competitive state rather than a near-tie. The toss covariate is informative here because a five-day pitch changes its nature and often deteriorates, so its magnitude is essentially format-dependent. The pairwise country-level home impact of eq.~(\ref{eq:define-home-impact}) reflects the conditions-driven, strongly opponent-specific nature of home advantage in Test cricket. For example, a spin-oriented side like Bangladesh or Sri Lanka touring a seam- or swing-oriented country like New Zealand or England poses a very different proposition from the reverse; its sparsity, in a closed $9$-team pool with an unbalanced calendar, is also what forces the shrinkage step.

\textbf{Extension to other formats and other sports.} The transfer to One Day Internationals (ODIs) and Twenty20 (T20s) is the most direct extension and needs several adjustments. The score set becomes binary and the draw layer can be dropped. The scale  $d$ must be recalibrated for a larger pool and a wider spread of strengths. The rating deviations will settle lower because teams play far more often, though the discounting will still matter for nations that play seldom. The toss impact operates based on pitch conditions, presence of dew, chasing advantage in day-night matches and not primarily through pitch deterioration, whereas the home advantage is weaker over a single day of play. The margin of victory is also reliably informative in the shorter formats and could be admitted as an additional covariate. Moreover, unlike Test cricket, rain-interrupted or shortened ODIs (and other 50-over matches, officially termed List A matches) and T20s often adhere to situation-specific modified targets and modified bowling and fielding restrictions based on the Duckworth-Lewis (DL Standard, DL Pro etc.), the Duckworth-Lewis-Stern (DLS) and the V Jayadevan (VJD) system, among others, all of which will then force a revised expected score model. Beyond cricket, the contextual component is again what must be redesigned: in association football, draws are frequent, so the three-outcome score set carries over, and the toss impact can be replaced by covariates such as rest days, travel fatigue factors, team strengths etc.; in basketball, there are no draws and the margin of victory is reliable, so a margin-based extension in the spirit of \cite{kovalchik2020extension} would combine naturally with the rating-deviation machinery retained here; in tennis, the playing surface is strongly player-specific, so the pairwise construction of eq.~(\ref{eq:define-home-impact}) is a closer template than a single global coefficient. In every case, the domain-independent components can be adopted unchanged, and it is that protocol, rather than the particular constants reported here, that we regard as the transferable contribution.

\section{Conclusion}\label{sec:conclusion}

This paper developed an enhanced, probabilistically grounded rating framework for Test cricket by adapting the Glicko rating system to the structure of the format. The methodological changes are (i) recalibration of the Elo/Glicko scaling factor (we find an optimal value $d=20$ for Test cricket, substantially smaller than the chess default $d=400$), (ii) incorporation of non-performance based factors, i.e. home-ground advantage and toss impact, into the expected-score formulation, with their statistical significance established by hypothesis testing, their additive form motivated from the Bradley-Terry construction and confirmed by a likelihood ratio test, and their relative weights estimated from data and shrunk on out-of-sample predictive loss, and (iii) a benchmarking of the framework against the standard Elo and unmodified Glicko systems. These modelling choices produce expected scores and rating updates that reflect the predictability of Test match outcomes.

The inclusion of home-ground advantage and toss effects addresses a factor that single-factor schemes ignore entirely, while the calibrated scaling ($d=20$) ensures that rating differences map to win probabilities appropriate to Test cricket's narrower rating dispersion and higher randomness. Estimating the contextual weights from data keeps the model parsimonious: home advantage is found to be worth about $13$ rating points and winning the toss about $8$ rating points, and the home and toss effects show no statistically significant interaction, so they enter the model additively.

The proposed model demonstrates strong predictive performance on the two completed World Test Championship cycles so far, viz.\ 2021-23 and 2023-25. Taking the team with the higher expected score as the predicted winner, the model correctly predicts $45$ of the $58$ non-drawn matches (approximately $77.6\%$ predictive accuracy) during WTC 2021-23. Across both completed cycles, it attains lower Brier score and log-loss than both the standard Elo and the unmodified Glicko systems while matching the better of their accuracies on decisive matches, so the gain is specifically one of probabilistic calibration.

The framework quantifies the uncertainty attached to the ratings at two levels. First, each team's rating carries an explicit rating deviation, initialised by the bootstrap standard errors of Section~\ref{sec:initialisation} and updated after every match, so the precision of every estimate is reported alongside its point estimate. Second, to determine whether end-of-cycle ratings are influenced by the particular order in which fixtures occurred, we construct a bootstrap distribution of each team's final rating over $B = 1000$ random permutations of the 70 matches in a cycle (Section~\ref{sec:resampling}). These distributions are tightly concentrated: the coefficient of variation averages 0.53\% in WTC 2021-23 and 0.57\% in WTC 2023-25, with per-team standard deviations of approximately half a rating point (between 0.42 and 0.61) throughout, and every team's chronological rating falls inside its $95\%$ bootstrap interval in both cycles, so the null of order-invariance is not rejected at any practically meaningful level. As set out in Section~\ref{sec:resampling}, we read this as establishing robustness to match ordering and not fairness with respect to the fixture list, which would require perturbing the schedule itself rather than the order of the matches it produced.

It is worth stating what has been demonstrated relative to the existing ICC procedure, particularly in view of the very high rank correlations of $0.979$ and $0.983$ between the two orderings. First, the framework supplies an expected score for each match, materially better calibrated than that of either baseline, on both the Brier score and the log-loss and in both cycles, at no cost in accuracy. The ICC scheme produces no such expected score or near-equivalent win probability, and therefore, cannot be scored on this criterion. Second, every rating carries an explicit, updated measure of its own precision, so a result against a well-established opponent and one against a rarely-seen opponent are no longer treated as equally informative, which is the deficiency of the ICC update identified in Section~\ref{sec:icc-ratings}. Third, the two contextual factors the ICC scheme omits are shown to be statistically significant and are priced in interpretable rating-point units, so their contribution is quantified rather than assumed. We have not demonstrated that the proposed system ranks the $9$ Test teams more accurately than the ICC scheme does: the two orderings differ by a single adjacent transposition in each cycle, and with only $9$ teams and no external criterion of true strength, no comparison on this data could settle which is closer to correct. Their near-coincidence is in fact reassuring, since it shows the calibration gains are not obtained by ranking teams implausibly. The contribution is therefore a probabilistically coherent, uncertainty-aware rating system that reproduces the accepted ordering while supplying the quantities the current procedure does not.

We deliberately restrict the model to the contextual factors whose significance we could establish on publicly available data, and do not incorporate further exogenous variables such as pitch conditions, weather, or player-level information. While such factors undoubtedly influence individual matches, their effects are difficult to quantify reliably at the team-season level and are only sparsely observed across the nine teams and two cycles studied here. Including them risks over-fitting and inflating apparent accuracy in-sample at the cost of out-of-sample generalisation, the failure mode the calibration design of Section~\ref{sec:adjust-400} is intended to avoid. We therefore treat them as a principled exclusion rather than an oversight.

Taken together, these results make the proposed framework a statistically principled, interpretable and order-robust alternative to the current Test rating procedure. Section~\ref{sec:generalisability} sets out which of its components are domain-independent and could be carried over to the limited-overs formats or to other sports, and which are calibrated to Test cricket and would have to be re-estimated first. Natural extensions include the incorporation of venue-specific and pitch-condition metrics as and when richer data permit their effects to be estimated without over-fitting, the admission of the margin of victory as a further covariate in formats where it is reliably informative, and a properly held-out validation of the tuning constants once a third World Test Championship cycle is complete.

\section*{Data availability statement}

The data used in this work have been obtained from two sources, both available for free: the schedule, results and match details of all the Test matches during our training window and the ICC WTC cycles from Cricinfo's Statsguru (link: \href{https://stats.cricinfo.com/ci/engine/stats/index.html}{https://stats.cricinfo.com/ci/engine/stats/index.html}) and the chronological published ICC men's Test team ratings from the ICC's website (link: \href{https://www.icc-cricket.com/rankings/team-rankings/mens/test}{https://www.icc-cricket.com/rankings/team-rankings/mens/test}). All the necessary code, in their reproducible form, will be made publicly available in the corresponding author's GitHub repository.


\bibliographystyle{apalike} 
\bibliography{references}

\newpage

\appendix

\section{Appendix}\label{sec:appendix}

\subsection{Prediction of Drawn Test Matches}\label{sec:draws}

The expected scores discussed in the paper address decisive results, but a non-negligible fraction of Test matches end in a draw. 19 out of the 150 matches, i.e.\ $12.67 \%$, ended up in draws during the training period, a share which increases slightly to $17.14 \%$ (12 out of 70 matches) during WTC 2021-23 and then drops to $5.71\%$ (4 out of 70 matches) during WTC 2023-25. Although in a match between teams $A$ and $B$, $E_A$ and $E_B$ denote $\mathbb{P}$(A wins the match) and $\mathbb{P}$(B wins the match) in some way, $(1 - E_A - E_B)$ does not alone necessarily determine the probability of $A$ and $B$ ending up in a drawn match. Note that $E_A$ and $E_B$ returning close values imply similar strengths of the teams competing in a match which in turn, also contributes to the possibility of the match leading to a draw. Thus, to include the effect of both these parameters in prediction of drawn Test matches, we consider, for $\alpha \in (0,1)$, a convex combination,

\begin{equation}
\label{eq:convex-comb-draws}
D_{\alpha, A, B} = \alpha (1 - E_A - E_B) + (1 - \alpha) |E_A - E_B|
\end{equation}

Bearing in mind the effect of exogenous variables such as rain-affected matches or other adverse weather conditions, flat pitches\footnote{Pitches suitable to batting where wickets do not fall frequently are often qualitatively termed as flat pitches.}, imposition of suspensions or draws due to political reasons, security concerns or any other unavoidable circumstances, one can argue that it is practically impossible to predict every drawn Test match. Our goal is to predict a significantly large number of drawn Test matches based on the convex combination of performance-based metrics.

\begin{table}[H]
    \centering
    \caption{Trade-off between choice of $\alpha$ and $q$-th quantile of $D_{\alpha, A, B}$ for prediction of drawn Test matches}
    \label{tab:choice-of-alpha-draws}
\begin{tabular}{| c | c c c c c c c c |}
\hline
    & \multicolumn{8}{c|}{Top \(100(1 - q) \%\)} \\
    \(\alpha\) & 35\% & 33\% & 30\% & 25\% & 20\% & 15\% & 10\% & 5\% \\
    \hline
    0.00 & 7 & 7 & 4 & 2 & 2 & 1 & 1 & 1 \\
    0.05 & 7 & 7 & 4 & 4 & 2 & 2 & 1 & 1 \\
    0.10 & 8 & 8 & 4 & 4 & 2 & 2 & 1 & 1 \\
    0.15 & 8 & 8 & 4 & 4 & 3 & 2 & 1 & 1 \\
    0.20 & 8 & 8 & 5 & 4 & 3 & 2 & 1 & 1 \\
    0.25 & 8 & 8 & 5 & 4 & 4 & 2 & 2 & 1 \\
    0.30 & 8 & 8 & 5 & 4 & 4 & 2 & 1 & 1 \\
    0.35 & 8 & 8 & 5 & 4 & 4 & 2 & 1 & 1 \\
    0.40 & 8 & 8 & 5 & 5 & 4 & 2 & 1 & 1 \\
    0.45 & 8 & 8 & 5 & 4 & 4 & 2 & 2 & 1 \\
    0.50 & 8 & 8 & 5 & 4 & 4 & 3 & 1 & 1 \\
    0.55 & 9 & 9 & 7 & 6 & 4 & 4 & 3 & 3 \\
    0.60 & 9 & 9 & 8 & 7 & 6 & 4 & 3 & 3 \\
    0.65 & 9 & 9 & 7 & 6 & 4 & 4 & 3 & 3 \\
    0.70 & 8 & 8 & 6 & 6 & 4 & 3 & 3 & 3 \\
    0.75 & 8 & 8 & 6 & 4 & 4 & 2 & 1 & 1 \\
    0.80 & 8 & 8 & 6 & 4 & 2 & 2 & 1 & 1 \\
    0.85 & 8 & 8 & 4 & 3 & 3 & 2 & 1 & 1 \\
    0.90 & 8 & 8 & 4 & 2 & 2 & 1 & 1 & 1 \\
    0.95 & 8 & 8 & 4 & 2 & 1 & 1 & 1 & 1 \\
    1.00 & 8 & 8 & 4 & 3 & 2 & 2 & 1 & 1 \\
    \hline
\end{tabular}
\end{table}

A well predicted drawn Test match is expected to produce a high value of $D_{\alpha, A, B}$ in (\ref{eq:convex-comb-draws}) for a certain choice of $\alpha \in (0,1)$. Equivalently, we might be interested in finding a choice of $\alpha$ for which a significantly large proportion of matches having high $D_{\alpha, A, B}$ values actually result in draws. We consider a trade-off between choices of $\alpha \in (0,1)$ and choices of a significantly large proportion, i.e.\ the $q$-th quantile of the $D_{\alpha, A, B}$ values of all the matches in a specific time period (see Table~\ref{tab:choice-of-alpha-draws}). From the table, the number of accurately predicted draws that fall in the top $33 \%$ and $35 \%$ are equal for any choice of $\alpha \in (0,1)$. For our test data of WTC 2021-23, $\alpha \in [0.55, 0.6]$ yield a high proportion ($9$ out of $12$) of accurately predicted drawn Test matches. A high proportion is maintained for the particular choice of $\alpha = 0.6$ for several choices of the quantile $q$. Based on our findings from the test data, we can consider $(0.6, 0.67)$ to be a sensible choice of $(\alpha, q)$.

\subsection{Additional tables}\label{sec:tables}

\textbf{Note:} Throughout this section, standard abbreviations have been used for each of the teams for ease of reference and to save space: AUS for Australia, BAN for Bangladesh, ENG for England, IND for India, NZ for New Zealand, PAK for Pakistan, SA for South Africa, SL for Sri Lanka and WI for West Indies.

\begin{table}[htbp]
    \centering
    \caption{Overview of ratings of chess players compared to Test cricket teams}
    \label{tab:chess-vs-cricket}
    \begin{adjustbox}{max width=\textwidth}
    \begin{tabular}{| l c | l c |}
\hline
Chess player categories & Glicko ratings & Test teams & Test ratings \\
\hline
Super Grandmasters & 2700+ & Australia & 124 \\
Most Grandmasters (GM) & 2500-2700 & India & 120 \\
Most International Masters (IM) \& some GM & 2400-2500 & England & 108 \\
Most FIDE Masters (FM) \& some IM & 2300-2400 & South Africa & 104 \\
FIDE Candidate Masters (CM) \& National Masters & 2200-2300 & New Zealand & 96 \\
Candidate Masters \& Experts & 2000-2200 & Sri Lanka & 83 \\
Class A, Category 1 & 1800-2000 & West Indies & 77 \\
Class B, Category 2 & 1600-1800 & Pakistan & 76 \\
Class C and below & below 1600 & Bangladesh & 66 \\
\hline
\end{tabular}
\end{adjustbox}
\end{table}

\begin{table}[htbp]
\centering
\caption{Comparison of loss functions across the scaling factor $d$ for expected-score calibration, each loss averaged across the two completed WTC cycles. Brier score, log-loss, MAE and ECE are to be minimised.}
\label{tab:scaling-factor}
\begin{tabular}{| c | c c c c |}
\hline
$d$ & Brier Score & Log-Loss & MAE & ECE \\
\hline
6 & 0.2070 & 0.7409 & 0.3543 & 0.1856 \\
10 & 0.1945 & 0.6535 & 0.3658 & 0.1502 \\
15 & 0.1875 & 0.6223 & 0.3780 & 0.1299 \\
\textbf{20} & \textbf{0.1856} & \textbf{0.6160} & 0.3875 & 0.1262 \\
25 & 0.1860 & 0.6174 & 0.3950 & 0.1158 \\
30 & 0.1874 & 0.6211 & 0.4009 & 0.0982 \\
40 & 0.1911 & 0.6299 & 0.4094 & 0.1049 \\
60 & 0.1974 & 0.6441 & 0.4194 & 0.1201 \\
85 & 0.2028 & 0.6554 & 0.4258 & 0.1263 \\
400 & 0.2168 & 0.6838 & 0.4391 & 0.1614 \\
\hline
\end{tabular}
\end{table}

\begin{table}[htbp]
\centering
\caption{Averaged Brier score over the grid of the scaling factor $d$ and the shrinkage factor $\lambda$, each entry averaged across the two completed WTC cycles.
$\lambda = 0$ switches the contextual weights off and therefore reproduces the Brier column of Table~\ref{tab:scaling-factor}. The minimum, 0.1837, is attained at $(d,\lambda) = (20, 0.1)$ and $(25, 0.2)$.}
\label{tab:dlambdagrid}
\begin{tabular}{| c | cccccccc |}
\hline
$d$ & $\lambda=0$ & $\lambda=0.05$ & $\lambda=0.1$ & $\lambda=0.2$ & $\lambda=0.3$ & $\lambda=0.5$ & $\lambda=0.75$ & $\lambda=1$ \\
\hline
6   & 0.2070 & 0.2069 & 0.2081 & 0.2127 & 0.2189 & 0.2301 & 0.2403 & 0.2506 \\
10  & 0.1945 & 0.1938 & 0.1942 & 0.1970 & 0.2017 & 0.2128 & 0.2249 & 0.2356 \\
15  & 0.1875 & 0.1864 & 0.1861 & 0.1876 & 0.1909 & 0.2001 & 0.2122 & 0.2231 \\
20  & 0.1856 & 0.1844 & \textbf{0.1837} & 0.1841 & 0.1862 & 0.1933 & 0.2040 & 0.2144 \\
25  & 0.1860 & 0.1848 & 0.1840 & \textbf{0.1837} & 0.1849 & 0.1900 & 0.1989 & 0.2082 \\
30  & 0.1874 & 0.1862 & 0.1854 & 0.1847 & 0.1851 & 0.1887 & 0.1958 & 0.2039 \\
40  & 0.1911 & 0.1900 & 0.1891 & 0.1880 & 0.1877 & 0.1890 & 0.1933 & 0.1991 \\
60  & 0.1974 & 0.1965 & 0.1958 & 0.1945 & 0.1937 & 0.1932 & 0.1943 & 0.1969 \\
85  & 0.2028 & 0.2021 & 0.2015 & 0.2004 & 0.1995 & 0.1983 & 0.1979 & 0.1985 \\
400 & 0.2168 & 0.2166 & 0.2164 & 0.2161 & 0.2158 & 0.2151 & 0.2144 & 0.2137 \\
\hline
\end{tabular}
\end{table}

\setlength{\LTcapwidth}{\textwidth}
\begin{longtable}{ccccccccccc}
\caption{Expected scores, updated ratings and rating deviations for every team in each match in WTC 2021-23, after imposition of home and away impacts and toss impact.}
\label{tab:full-list-wtc2021-23}
\\
\hline
\textbf{Date} & \textbf{A} & \textbf{B} & \textbf{Toss} & \textbf{$E_A$} & \textbf{$E_B$} & \textbf{Winner} & \textbf{$R_A$} & \textbf{$R_B$} & \textbf{$RD_A$} & \textbf{$RD_B$} \\
\hline
\endfirsthead

\hline
\textbf{Date} & \textbf{A} & \textbf{B} & \textbf{Toss} & \textbf{$E_A$} & \textbf{$E_B$} & \textbf{Winner} & \textbf{$R_A$} & \textbf{$R_B$} & \textbf{$RD_A$} & \textbf{$RD_B$} \\
\hline
\endhead
4 Aug 2021 & ENG & IND & ENG & 0.49 & 0.51 & Draw & 108.01 & 119.99 & 8.43 & 8.41 \\
12 Aug 2021 & ENG & IND & ENG & 0.49 & 0.51 & IND & 107.36 & 120.64 & 6.30 & 6.30 \\
12 Aug 2021 & WI & PAK & WI & 0.48 & 0.51 & WI & 77.67 & 75.28 & 8.35 & 8.44 \\
20 Aug 2021 & WI & PAK & WI & 0.47 & 0.51 & PAK & 77.05 & 75.93 & 6.28 & 6.29 \\
25 Aug 2021 & ENG & IND & IND & 0.48 & 0.52 & ENG & 108.04 & 119.95 & 4.75 & 4.75 \\
2 Sep 2021 & ENG & IND & ENG & 0.48 & 0.52 & IND & 107.42 & 120.58 & 3.63 & 3.63 \\
21 Nov 2021 & SL & WI & SL & 0.53 & 0.49 & SL & 83.74 & 76.54 & 5.66 & 5.36 \\
25 Nov 2021 & IND & NZ & IND & 0.52 & 0.36 & Draw & 120.57 & 96.28 & 3.60 & 4.58 \\
26 Nov 2021 & BAN & PAK & BAN & 0.48 & 0.51 & PAK & 65.16 & 76.31 & 6.38 & 5.81 \\
29 Nov 2021 & SL & WI & SL & 0.54 & 0.48 & SL & 84.37 & 75.93 & 4.19 & 4.15 \\
3 Dec 2021 & IND & NZ & IND & 0.61 & 0.36 & IND & 121.01 & 95.75 & 3.02 & 3.25 \\
4 Dec 2021 & BAN & PAK & PAK & 0.45 & 0.55 & PAK & 64.54 & 76.86 & 4.64 & 4.56 \\
8 Dec 2021 & AUS & ENG & ENG & 0.65 & 0.46 & AUS & 124.70 & 107.22 & 4.48 & 3.55 \\
16 Dec 2021 & AUS & ENG & AUS & 0.65 & 0.37 & AUS & 125.21 & 106.81 & 3.21 & 2.98 \\
26 Dec 2021 & AUS & ENG & AUS & 0.67 & 0.33 & AUS & 125.65 & 106.41 & 2.53 & 2.46 \\
26 Dec 2021 & SA & IND & IND & 0.38 & 0.51 & IND & 103.23 & 121.10 & 3.97 & 3.01 \\
1 Jan 2022 & NZ & BAN & BAN & 0.69 & 0.27 & BAN & 95.03 & 65.70 & 2.85 & 3.27 \\
3 Jan 2022 & SA & IND & IND & 0.37 & 0.59 & SA & 104.14 & 120.47 & 2.82 & 2.56 \\
5 Jan 2022 & AUS & ENG & AUS & 0.70 & 0.29 & Draw & 125.40 & 106.66 & 2.08 & 2.06 \\
9 Jan 2022 & NZ & BAN & BAN & 0.73 & 0.26 & NZ & 95.35 & 65.34 & 2.43 & 2.59 \\
11 Jan 2022 & SA & IND & IND & 0.37 & 0.61 & SA & 104.95 & 119.77 & 2.22 & 2.12 \\
14 Jan 2022 & AUS & ENG & ENG & 0.73 & 0.28 & AUS & 125.72 & 106.34 & 1.78 & 1.76 \\
17 Feb 2022 & NZ & SA & NZ & 0.43 & 0.55 & NZ & 96.05 & 104.34 & 1.93 & 1.85 \\
25 Feb 2022 & NZ & SA & SA & 0.42 & 0.59 & SA & 95.57 & 104.78 & 1.61 & 1.57 \\
4 Mar 2022 & IND & SL & IND & 0.69 & 0.21 & IND & 120.01 & 83.99 & 1.98 & 2.82 \\
4 Mar 2022 & PAK & AUS & PAK & 0.21 & 0.69 & Draw & 77.44 & 125.60 & 2.75 & 1.70 \\
8 Mar 2022 & WI & ENG & ENG & 0.29 & 0.63 & Draw & 76.30 & 106.25 & 2.47 & 1.66 \\
12 Mar 2022 & PAK & AUS & AUS & 0.15 & 0.75 & Draw & 78.00 & 125.38 & 2.23 & 1.58 \\
12 Mar 2022 & IND & SL & IND & 0.75 & 0.19 & IND & 120.25 & 83.69 & 1.80 & 2.25 \\
16 Mar 2022 & WI & ENG & ENG & 0.29 & 0.69 & Draw & 76.60 & 106.08 & 1.90 & 1.51 \\
21 Mar 2022 & PAK & AUS & AUS & 0.15 & 0.78 & AUS & 77.78 & 125.58 & 1.91 & 1.46 \\
24 Mar 2022 & WI & ENG & WI & 0.27 & 0.70 & WI & 77.50 & 105.42 & 1.60 & 1.36 \\
31 Mar 2022 & SA & BAN & BAN & 0.74 & 0.20 & SA & 104.99 & 65.02 & 1.46 & 2.03 \\
8 Apr 2022 & SA & BAN & SA & 0.80 & 0.16 & SA & 105.17 & 64.81 & 1.36 & 1.76 \\
15 May 2022 & BAN & SL & SL & 0.29 & 0.72 & Draw & 65.02 & 83.40 & 1.58 & 1.83 \\
23 May 2022 & BAN & SL & BAN & 0.32 & 0.70 & SL & 64.70 & 83.74 & 1.40 & 1.55 \\
2 Jun 2022 & ENG & NZ & NZ & 0.53 & 0.47 & ENG & 105.85 & 95.06 & 1.22 & 1.35 \\
10 Jun 2022 & ENG & NZ & ENG & 0.55 & 0.45 & ENG & 106.25 & 94.62 & 1.09 & 1.18 \\
16 Jun 2022 & WI & BAN & WI & 0.72 & 0.25 & WI & 77.80 & 64.46 & 1.39 & 1.27 \\
23 Jun 2022 & ENG & NZ & NZ & 0.55 & 0.45 & ENG & 106.63 & 94.21 & 1.00 & 1.06 \\
24 Jun 2022 & WI & BAN & WI & 0.73 & 0.24 & WI & 78.07 & 64.24 & 1.24 & 1.17 \\
29 Jun 2022 & SL & AUS & SL & 0.17 & 0.84 & AUS & 83.55 & 125.75 & 1.41 & 1.35 \\
1 Jul 2022 & ENG & IND & ENG & 0.45 & 0.56 & ENG & 106.99 & 119.56 & 0.94 & 1.41 \\
8 Jul 2022 & SL & AUS & AUS & 0.15 & 0.83 & SL & 84.45 & 124.92 & 1.31 & 1.25 \\
16 Jul 2022 & SL & PAK & SL & 0.57 & 0.39 & PAK & 83.98 & 78.54 & 1.19 & 1.52 \\
24 Jul 2022 & SL & PAK & SL & 0.57 & 0.39 & SL & 84.34 & 78.12 & 1.09 & 1.29 \\
17 Aug 2022 & ENG & SA & SA & 0.58 & 0.44 & SA & 106.59 & 105.76 & 0.88 & 1.17 \\
25 Aug 2022 & ENG & SA & SA & 0.57 & 0.45 & ENG & 106.88 & 105.34 & 0.83 & 1.04 \\
8 Sep 2022 & ENG & SA & ENG & 0.59 & 0.38 & ENG & 107.16 & 105.01 & 0.78 & 0.94 \\
30 Nov 2022 & AUS & WI & AUS & 0.87 & 0.13 & AUS & 125.05 & 77.94 & 1.18 & 1.17 \\
1 Dec 2022 & PAK & ENG & ENG & 0.17 & 0.77 & ENG & 77.93 & 107.30 & 1.18 & 0.75 \\
8 Dec 2022 & AUS & WI & AUS & 0.87 & 0.13 & AUS & 125.17 & 77.83 & 1.12 & 1.12 \\
9 Dec 2022 & PAK & ENG & ENG & 0.17 & 0.78 & ENG & 77.75 & 107.44 & 1.09 & 0.73 \\
14 Dec 2022 & BAN & IND & IND & 0.09 & 0.91 & IND & 64.16 & 119.66 & 1.13 & 1.34 \\
17 Dec 2022 & AUS & SA & AUS & 0.70 & 0.28 & AUS & 125.45 & 104.80 & 1.02 & 0.89 \\
17 Dec 2022 & PAK & ENG & PAK & 0.24 & 0.78 & ENG & 77.53 & 107.57 & 1.00 & 0.71 \\
22 Dec 2022 & BAN & IND & BAN & 0.11 & 0.91 & IND & 64.07 & 119.75 & 1.09 & 1.28 \\
26 Dec 2022 & AUS & SA & AUS & 0.70 & 0.27 & AUS & 125.70 & 104.61 & 0.94 & 0.84 \\
26 Dec 2022 & PAK & NZ & PAK & 0.33 & 0.71 & Draw & 77.67 & 94.04 & 0.93 & 0.97 \\
2 Jan 2023 & PAK & NZ & NZ & 0.25 & 0.72 & Draw & 77.86 & 93.86 & 0.87 & 0.91 \\
4 Jan 2023 & AUS & SA & AUS & 0.71 & 0.26 & Draw & 125.53 & 104.78 & 0.88 & 0.80 \\
9 Feb 2023 & IND & AUS & AUS & 0.43 & 0.55 & IND & 120.32 & 125.16 & 1.11 & 0.83 \\
17 Feb 2023 & IND & AUS & AUS & 0.44 & 0.54 & IND & 120.83 & 124.80 & 0.99 & 0.78 \\
28 Feb 2023 & SA & WI & SA & 0.78 & 0.24 & SA & 104.92 & 77.60 & 0.77 & 1.02 \\
1 Mar 2023 & IND & AUS & IND & 0.46 & 0.56 & AUS & 120.45 & 125.09 & 0.90 & 0.74 \\
8 Mar 2023 & SA & WI & SA & 0.78 & 0.23 & SA & 105.06 & 77.40 & 0.74 & 0.95 \\
9 Mar 2023 & NZ & SL & NZ & 0.61 & 0.38 & NZ & 94.15 & 84.01 & 0.85 & 0.98 \\
9 Mar 2023 & IND & AUS & AUS & 0.44 & 0.54 & Draw & 120.49 & 125.06 & 0.83 & 0.70 \\
17 Mar 2023 & NZ & SL & SL & 0.60 & 0.41 & NZ & 94.42 & 83.67 & 0.80 & 0.90 \\
7 Jun 2023 & AUS & IND & IND & 0.53 & 0.49 & AUS & 125.35 & 120.14 & 0.67 & 0.78 \\
\hline
\end{longtable}

\setlength{\LTcapwidth}{\textwidth}
\begin{longtable}{ccccccccccc}
\caption{Expected scores, updated ratings and rating deviations for every team in each match in WTC 2023-25, after imposition of home and away impacts and toss impact.}
\label{tab:full-list-wtc2023-25}
\\
\hline
\textbf{Date} & \textbf{A} & \textbf{B} & \textbf{Toss} & \textbf{$E_A$} & \textbf{$E_B$} & \textbf{Winner} & \textbf{$R_A$} & \textbf{$R_B$} & \textbf{$RD_A$} & \textbf{$RD_B$} \\
\hline
\endfirsthead

\hline
\textbf{Date} & \textbf{A} & \textbf{B} & \textbf{Toss} & \textbf{$E_A$} & \textbf{$E_B$} & \textbf{Winner} & \textbf{$R_A$} & \textbf{$R_B$} & \textbf{$RD_A$} & \textbf{$RD_B$} \\
\hline
\endhead
16 Jun 2023 & ENG & AUS & ENG & 0.48 & 0.54 & AUS & 107.28 & 125.57 & 0.68 & 0.64 \\
28 Jun 2023 & ENG & AUS & ENG & 0.46 & 0.55 & AUS & 107.00 & 125.78 & 0.69 & 0.65 \\
6 Jul 2023 & ENG & AUS & ENG & 0.44 & 0.57 & ENG & 107.23 & 125.54 & 0.69 & 0.65 \\
12 Jul 2023 & WI & IND & WI & 0.39 & 0.61 & IND & 77.20 & 120.32 & 0.91 & 0.75 \\
16 Jul 2023 & SL & PAK & SL & 0.51 & 0.47 & PAK & 83.45 & 78.15 & 0.86 & 0.84 \\
19 Jul 2023 & ENG & AUS & ENG & 0.43 & 0.58 & Draw & 107.22 & 125.53 & 0.65 & 0.61 \\
20 Jul 2023 & WI & IND & WI & 0.35 & 0.65 & Draw & 77.22 & 119.88 & 0.91 & 0.75 \\
24 Jul 2023 & SL & PAK & SL & 0.52 & 0.47 & PAK & 83.22 & 78.45 & 0.86 & 0.84 \\
27 Jul 2023 & ENG & AUS & AUS & 0.41 & 0.59 & ENG & 107.47 & 125.29 & 0.66 & 0.61 \\
28 Nov 2023 & BAN & NZ & BAN & 0.48 & 0.55 & BAN & 64.37 & 94.19 & 1.05 & 0.77 \\
6 Dec 2023 & BAN & NZ & BAN & 0.46 & 0.55 & NZ & 64.17 & 94.45 & 1.04 & 0.77 \\
14 Dec 2023 & AUS & PAK & AUS & 0.66 & 0.22 & AUS & 125.46 & 78.33 & 0.61 & 0.84 \\
26 Dec 2023 & AUS & PAK & PAK & 0.74 & 0.25 & AUS & 125.60 & 78.20 & 0.57 & 0.79 \\
26 Dec 2023 & SA & IND & SA & 0.46 & 0.51 & SA & 106.39 & 118.26 & 0.71 & 0.75 \\
3 Jan 2024 & AUS & PAK & PAK & 0.79 & 0.23 & AUS & 125.72 & 78.08 & 0.57 & 0.79 \\
3 Jan 2024 & SA & IND & SA & 0.46 & 0.54 & IND & 106.19 & 118.46 & 0.71 & 0.70 \\
17 Jan 2024 & AUS & WI & AUS & 0.66 & 0.18 & AUS & 125.89 & 77.11 & 0.57 & 0.90 \\
25 Jan 2024 & AUS & WI & WI & 0.77 & 0.16 & WI & 125.58 & 77.44 & 0.54 & 0.85 \\
25 Jan 2024 & IND & ENG & ENG & 0.61 & 0.42 & ENG & 116.54 & 107.71 & 0.70 & 0.66 \\
2 Feb 2024 & IND & ENG & IND & 0.62 & 0.39 & IND & 116.71 & 107.46 & 0.70 & 0.62 \\
4 Feb 2024 & NZ & SA & SA & 0.45 & 0.53 & NZ & 94.75 & 105.96 & 0.78 & 0.71 \\
13 Feb 2024 & NZ & SA & SA & 0.45 & 0.55 & NZ & 95.06 & 105.73 & 0.73 & 0.67 \\
15 Feb 2024 & IND & ENG & IND & 0.64 & 0.36 & IND & 116.88 & 107.23 & 0.66 & 0.62 \\
23 Feb 2024 & IND & ENG & ENG & 0.66 & 0.35 & IND & 117.04 & 107.01 & 0.65 & 0.62 \\
29 Feb 2024 & NZ & AUS & NZ & 0.27 & 0.66 & AUS & 94.92 & 125.75 & 0.73 & 0.54 \\
7 Mar 2024 & IND & ENG & ENG & 0.68 & 0.33 & IND & 117.19 & 106.80 & 0.65 & 0.59 \\
8 Mar 2024 & NZ & AUS & AUS & 0.25 & 0.70 & AUS & 94.80 & 125.91 & 0.73 & 0.53 \\
22 Mar 2024 & BAN & SL & BAN & 0.42 & 0.55 & SL & 63.99 & 83.42 & 1.02 & 0.86 \\
30 Mar 2024 & BAN & SL & SL & 0.38 & 0.61 & SL & 63.81 & 83.59 & 0.97 & 0.81 \\
10 Jul 2024 & ENG & WI & ENG & 0.72 & 0.22 & ENG & 106.93 & 77.31 & 0.59 & 0.85 \\
18 Jul 2024 & ENG & WI & WI & 0.75 & 0.20 & ENG & 107.06 & 77.19 & 0.59 & 0.84 \\
26 Jul 2024 & ENG & WI & WI & 0.77 & 0.19 & ENG & 107.18 & 77.07 & 0.56 & 0.80 \\
7 Aug 2024 & WI & SA & SA & 0.36 & 0.76 & Draw & 77.09 & 105.98 & 0.78 & 0.67 \\
15 Aug 2024 & WI & SA & SA & 0.32 & 0.76 & SA & 76.92 & 106.59 & 0.77 & 0.67 \\
21 Aug 2024 & PAK & BAN & BAN & 0.62 & 0.32 & BAN & 77.83 & 64.20 & 0.79 & 0.96 \\
21 Aug 2024 & ENG & SL & SL & 0.62 & 0.28 & ENG & 107.35 & 83.45 & 0.56 & 0.81 \\
29 Aug 2024 & ENG & SL & SL & 0.69 & 0.27 & ENG & 107.50 & 83.31 & 0.55 & 0.80 \\
30 Aug 2024 & PAK & BAN & BAN & 0.65 & 0.32 & BAN & 77.56 & 64.58 & 0.75 & 0.94 \\
6 Sep 2024 & ENG & SL & SL & 0.71 & 0.26 & SL & 107.09 & 83.61 & 0.53 & 0.76 \\
18 Sep 2024 & SL & NZ & SL & 0.42 & 0.61 & SL & 83.86 & 94.54 & 0.75 & 0.70 \\
19 Sep 2024 & IND & BAN & BAN & 0.91 & 0.07 & IND & 117.26 & 64.52 & 0.61 & 0.90 \\
26 Sep 2024 & SL & NZ & SL & 0.42 & 0.60 & SL & 84.10 & 94.29 & 0.74 & 0.69 \\
27 Sep 2024 & IND & BAN & IND & 0.92 & 0.05 & IND & 117.33 & 64.46 & 0.61 & 0.88 \\
7 Oct 2024 & PAK & ENG & PAK & 0.25 & 0.74 & ENG & 77.43 & 107.22 & 0.74 & 0.52 \\
15 Oct 2024 & PAK & ENG & PAK & 0.24 & 0.76 & PAK & 77.84 & 106.77 & 0.74 & 0.52 \\
16 Oct 2024 & IND & NZ & IND & 0.76 & 0.24 & NZ & 115.33 & 94.70 & 0.60 & 0.68 \\
21 Oct 2024 & BAN & SA & BAN & 0.22 & 0.79 & SA & 64.34 & 107.13 & 0.86 & 0.63 \\
24 Oct 2024 & IND & NZ & NZ & 0.75 & 0.25 & NZ & 113.33 & 95.10 & 0.57 & 0.66 \\
24 Oct 2024 & PAK & ENG & ENG & 0.18 & 0.77 & PAK & 78.27 & 106.32 & 0.71 & 0.50 \\
29 Oct 2024 & BAN & SA & SA & 0.16 & 0.83 & SA & 64.25 & 107.57 & 0.82 & 0.63 \\
1 Nov 2024 & IND & NZ & NZ & 0.74 & 0.26 & NZ & 111.33 & 95.50 & 0.56 & 0.65 \\
22 Nov 2024 & AUS & IND & IND & 0.55 & 0.48 & IND & 125.67 & 111.55 & 0.51 & 0.55 \\
22 Nov 2024 & WI & BAN & BAN & 0.74 & 0.29 & WI & 77.05 & 64.11 & 0.74 & 0.80 \\
27 Nov 2024 & SA & SL & SL & 0.68 & 0.33 & SA & 108.36 & 83.94 & 0.62 & 0.71 \\
28 Nov 2024 & NZ & ENG & ENG & 0.37 & 0.62 & ENG & 95.33 & 106.50 & 0.64 & 0.49 \\
30 Nov 2024 & WI & BAN & BAN & 0.76 & 0.28 & BAN & 76.71 & 64.51 & 0.72 & 0.77 \\
5 Dec 2024 & SA & SL & SA & 0.73 & 0.28 & SA & 109.03 & 83.80 & 0.60 & 0.70 \\
6 Dec 2024 & NZ & ENG & NZ & 0.38 & 0.62 & ENG & 95.16 & 106.67 & 0.62 & 0.48 \\
6 Dec 2024 & AUS & IND & IND & 0.54 & 0.48 & AUS & 125.89 & 110.03 & 0.50 & 0.52 \\
14 Dec 2024 & NZ & ENG & ENG & 0.36 & 0.64 & NZ & 95.51 & 106.29 & 0.61 & 0.47 \\
14 Dec 2024 & AUS & IND & IND & 0.55 & 0.47 & Draw & 125.89 & 109.62 & 0.49 & 0.51 \\
26 Dec 2024 & AUS & IND & AUS & 0.52 & 0.46 & AUS & 126.11 & 108.16 & 0.47 & 0.50 \\
26 Dec 2024 & SA & PAK & SA & 0.74 & 0.24 & SA & 109.69 & 78.15 & 0.59 & 0.70 \\
3 Jan 2025 & AUS & IND & IND & 0.56 & 0.47 & AUS & 126.32 & 106.69 & 0.47 & 0.48 \\
3 Jan 2025 & SA & PAK & SA & 0.75 & 0.23 & SA & 110.33 & 78.03 & 0.58 & 0.68 \\
17 Jan 2025 & PAK & WI & PAK & 0.57 & 0.50 & PAK & 78.27 & 76.46 & 0.66 & 0.70 \\
25 Jan 2025 & PAK & WI & WI & 0.47 & 0.46 & WI & 78.07 & 76.69 & 0.65 & 0.68 \\
29 Jan 2025 & SL & AUS & AUS & 0.12 & 0.86 & AUS & 83.72 & 126.42 & 0.68 & 0.46 \\
6 Feb 2025 & SL & AUS & SL & 0.13 & 0.88 & AUS & 83.63 & 126.50 & 0.66 & 0.44 \\
11 Jun 2025 & SA & AUS & SA & 0.30 & 0.73 & SA & 112.02 & 126.21 & 0.56 & 0.43 \\
\hline
\end{longtable}

\end{document}